\documentclass[a4paper,10pt]{article}

\usepackage[utf8]{inputenc}

\usepackage{authblk}

\usepackage{ifthen,ifpdf}

\ifpdf
  \usepackage{hyperref}
  \hypersetup{colorlinks=false,allcolors=blue}
  \usepackage{hypcap}
  \pdfpagewidth=\paperwidth
  \pdfpageheight=\paperheight
\fi

\usepackage[utf8]{inputenc}
\usepackage[T1]{fontenc}
\usepackage{amssymb}
\usepackage{amsmath, amsthm}
\usepackage{mathtools}
\usepackage{numprint}
\usepackage{enumerate}
\usepackage{afterpage}
\usepackage{tabularx}
\usepackage{dsfont}
\usepackage{graphicx}

\usepackage{algorithm}
\usepackage{algpseudocode}
\usepackage{caption}
\usepackage[labelformat=simple]{subcaption}

\usepackage{color}
\usepackage{units}
\usepackage{array, multirow}
\usepackage{booktabs}

\newcolumntype{M}[1]{>{\vspace{3pt}\raggedleft\arraybackslash}m{#1}}

\usepackage{siunitx}
\usepackage{tikz}
\usepackage{pgfplots}
\usepackage{pgfplotstable}
\usepgfplotslibrary{units}
\pgfplotsset{compat=1.9}
\pgfplotstableset{col sep=comma}

\usepackage{cite}

\usepackage{anysize} 
\marginsize{2.5cm}{2.5cm}{2cm}{2cm}

\usepackage{wrapfig}

\theoremstyle{plain}

\theoremstyle{remark}

\numberwithin{equation}{section}

\newcommand{\R}{\mathds{R}}

\def\Xint#1{\mathchoice
   {\XXint\displaystyle\textstyle{#1}}%
   {\XXint\textstyle\scriptstyle{#1}}%
   {\XXint\scriptstyle\scriptscriptstyle{#1}}%
   {\XXint\scriptscriptstyle\scriptscriptstyle{#1}}%
   \!\int}
\def\XXint#1#2#3{{\setbox0=\hbox{$#1{#2#3}{\int}$}
     \vcenter{\hbox{$#2#3$}}\kern-.5\wd0}}

\def\dashint{\Xint-}

\DeclareMathOperator{\Id}{\textrm{Id}}

\newcommand{\review}[1]{{\color{black}{#1}}} 

\newcommand{\mean}[1]{ \left\langle {#1} \right\rangle}
\newcommand{\var}[1]{ \textrm{Var}\!\left( {#1} \right)}

\title{{Representative volume elements for matrix-inclusion composites - a computational study on periodizing the ensemble}}

\author[1,*]{Matti Schneider}
\author[2]{Marc Josien}
\author[3]{Felix Otto}

\affil[1]{Karlsruhe Institute of Technology (KIT), Institute of Engineering Mechanics}
\affil[2]{CEA, DES, IRESNE, DEC, Cadarache, F-13108, Saint-Paul-Lez-Durance, France}
\affil[3]{Max-Planck Institute for Mathematics in the Sciences (MPI MiS)}
\affil[*]{correspondence to: \texttt{matti.schneider@kit.edu}}

\date{\today}

\begin{document}

\maketitle 

\begin{abstract}
\noindent We investigate {volume-element} sampling strategies {for the stochastic} homogenization {of particle-reinforced composites} and show, via computational {experiments}, that an improper treatment of particles intersecting the boundary of the computational cell may {affect} the accuracy of the computed effective properties. Motivated by {recent} results on a superior convergence rate of the systematic error for {periodized ensembles compared to taking snapshots of ensembles}, we conduct computational experiments for microstructures with circular, spherical and cylindrical inclusions and monitor the systematic errors in the effective thermal conductivity for snapshots of ensembles compared to working with microstructures sampled from periodized ensembles.\\
We observe that the standard deviation {of the apparent properties computed on microstructures sampled from} the periodized ensembles decays at the scaling expected from the central limit theorem. In contrast, the standard deviation for the {snapshot} ensembles shows an inferior decay rate at high filler content. The latter effect is caused by additional long-range correlations that necessarily appear in particle-reinforced composites at high, industrially relevant, volume fractions. Periodized ensembles, however, appear to be {less} affected by these correlations.\\
Our findings provide guidelines for working with digital volume images of material microstructures and the design of representative volume elements for computational homogenization.
\end{abstract}

%

\newpage

\section{Introduction}
\label{sec:intro}

\subsection{State of the art}

{The concept of representative volume element (RVE) plays a central role for predicting the effective properties of random heterogeneous materials.} Originally, Hill~\cite{HillRVE} defined an RVE as a bounded domain {that} is statistically typical of the mixture and sufficiently large to render the effects of artificially imposed boundary conditions negligible. Drugan \& Willis~\cite{DruganWillis} improved upon the original concept by \review{relaxing} the previous definition, \review{that} requires the RVE to be typical for \emph{all} statistics of interest. Rather, they {focused on the material properties and} defined an RVE as a bounded domain whose apparent properties are sufficiently close to the effective properties of the infinitely large medium. This shift in paradigm{, which introduced the infinite-volume limit as a suitable reference quantity,} enabled much smaller RVEs to be used, with obvious advantages in terms of computational complexity. Kanit et al.~\cite{KanitRVE} proposed a more quantitative definition of the RVE based on {the statistics of the apparent properties on cells of varying, but finite, size}. They introduced the notions of bias and dispersion (corresponding to the systematic and the random error, respectively, to be discussed below) for the apparent properties, and provided theoretical arguments {as well as} empirical data \review{in support of} using several realizations of the {ensemble} on smaller volumes to arrive at the same accuracy as for a single "large" RVE. Improvements in computational power and algorithmic advances gave significant impetus to modern multiscale methods in engineering, also dealing with nonlinear and time-dependent problems, see Matou\v{s} et al.~\cite{MatousSummary} for a recent overview.\\
{In general, one may distinguish two succinctly different strategies for acquiring the microstructures necessary for multiscale methods in engineering. On the one hand, synthetic microstructures may be generated based on statistical data available on the material under consideration, see Bargmann et al.~\cite{BargmannReview} for a recent overview. On the other hand,} digital volume images of {real} microstructures~\cite{LandisKeane,GuvenCinar} {may be obtained directly{, serving as what we call snapshots of the random material}. Both approaches complement each other, and may be used cooperatively. Indeed, synthetic microstructures offer full control over the ensemble, and thorough RVE studies can be conducted. However, the relation of these synthetic structures to their real counterparts it is not always apparent. In contrast, {snapshots} are real, but their acquisition is typically costly, which makes it hard to study their representativity and to quantify the underlying uncertainty.}\\
Parallel to developments in engineering, homogenization theory was established as a rigorous foundation for the up-scaling of random heterogeneous media. Building upon previous work in the periodic setting~\cite{PeriodicHomogenization,PeriodicHomogenization2,PeriodicHomogenization3}, Varadhan \& Papanicolaou~\cite{PapanicolaouVaradhan} and Kozlov~\cite{Kozlov} established qualitative results for stochastic homogenization for stationary and ergodic {ensembles of} coefficient {fields}. In this context and informally speaking, stationarity means that the statistical properties of the ensembles are translation-invariant, whereas ergodicity means that {the microstructure becomes statistically independent over distances tending to infinity}.\\
Classically, the effective properties in stochastic homogenization are defined by {spatially} averaging so-called corrector fields which are defined on the {whole} space. The RVE methodology, i.e., working with cells of {large but} finite {size}, is imperative from a computational point of view. {The error of the apparent property associated to single finite-sized cell compared to the true effective properties of stochastic homogenization naturally decomposes into two parts~\cite[Eq. (13)]{GloriaNeukammOtto2015}. The random error quantifies the fluctuation of the apparent properties on cells of fixed size, whereas the systematic error measures the deviation of the {expectation of the} apparent property on cells of fixed size from the true effective property. These two error contributions} correspond to the bias and dispersion considered by Kanit et al.~\cite{KanitRVE}. Theoretical guarantees {that the systematic as well as the random error converge to zero as the cell size goes to infinity} were provided by Bourgeat \& Piatnitski~\cite{BourgeatPiatnitski} and Owhadi~\cite{Owhadi2003} for Dirichlet, Neumann and periodic boundary conditions imposed on the boundary of the considered cells.\\
These results hold in the general setting of stationary and ergodic random media. Thus, they may only provide \emph{qualitative} convergence results, i.e., convergence {of the random and the systematic error to zero}, without rates. To obtain specific convergence rates, the {correlation} properties of the coefficient field {need} to be quantified, {see Section \ref{sec:theory} below for a discussion}.\\
{Theoretical works also shed some light on the two different approaches for obtaining microstructures discussed earlier. For the mathematical analysis, the full ensemble of microstructures needs to be given, and its stochastic properties are encoded by the underlying correlation structure. Acquiring {snapshots} corresponds to drawing a specific sample of the ensemble, and restricting the microstructure to a cell of finite size. This approach is straightforward, and is {tacitly} assumed in the qualitative convergence results~\cite{BourgeatPiatnitski, Owhadi2003}. Generating synthetic microstructures is more subtle, as those are typically generated on a rectangular cell with periodic boundary conditions, i.e., on a torus, mathematically speaking. Thus, these generated elements are more than simple restrictions of the infinite-space ensemble, but {require the underlying {"construction plan"} to be known~{\cite[Sec. 3.2.3]{Egloffe2015}}{, for instance via pushing forward a specific probability distribution by a (deterministic) mapping}. With this more fundamental knowledge at hand, it is natural to draw} \review{samples from} suitably periodized ensembles for the finitely sized cells. In practical engineering terms, microstructure-generation algorithms are - in principle - deterministic, but some of its input parameters are drawn from probability distributions. Combining the random input and the deterministic algorithm gives rise to the {ensemble}, and drawing samples from the periodized {ensemble} corresponds to working on the torus instead of a box in Euclidean space, for instance when computing distances or applying filters.\\
Over the years, in terms of theoretical reasoning {as well as} numerical evidence, it {has been} realized that the {snapshot} {strategy} may introduce non-negligible boundary effects compared to the periodization approach. {This observation surfaced in numerical multiscale methods~\cite{HouWu1997,HMMAnalysis}, where the finite-element ansatz functions were modified in order to account for heterogeneity on a sub-element scale by solving a corrector-type equation with suitable boundary conditions. In the setting of periodic homogenization it was realized that working on {cubic} volume elements with an edge length $L$ that is not an integer multiple of the period leads to an error in the effective properties that is of order $1/L$. 
{To reduce this so-called resonance error, oversampling/filtering techniques~\cite{HouWu1997,Blanc2010} were introduced.} {Alternative} approaches screen the boundary effect by modifying the cell problem, for instance by adding a "massive", zero-th order term to the elliptic operator defining the corrector problem~\cite[Thm. 1]{gloria2011reduction}. In a similar spirit, {screening} strategies based on parabolic~\cite{FancyStuff2,FancyStuff1} or hyperbolic~\cite{Arjmand2016} versions of the corrector problem {have been} considered.\\
In the context of stochastic homogenization, the {snapshot} approach also gives rise to a boundary-layer error, just as for periodic homogenization with non-matching period.
In the stochastic setting, Egloffe et al.~\cite[Eq. (3.4)]{Egloffe2015} expected the systematic error to decay with the rate $1/L$, as for the periodic case. Recently, Clozeau et al.~\cite{CloJoOtXu_2020} provided a theoretical argument for the $1/L$-scaling, at least in the asymptotic setting of vanishing material contrast. {Because of the slow decay of the random boundary layer, filtering techniques {are expected to} have no effect on the scaling $1/L$~\cite[Sec. 4]{gloria2008windows}. However, screening by a massive term, in conjunction with extrapolation in the massive parameter, has been shown to reduce the systematic error to $1/L^d$~\cite[Thm. 2]{GloriaNeukammOtto2015}. Based on screening and extrapolation, Mourrat~\cite[Prop.\ 1.1 \& Th.\ 1.2]{FancyStuff1} devised a numerical algorithm that, on the basis of a {snapshot}, extracts the effective behavior up to the optimal total{-}error {rate} ${1/L^\frac{d}{2}}$ at the
cost of {on the order of} ${L^d}$ operations.}\\
{Hence, the} most effective improvements in the decay of the boundary-layer error rest upon a modification of the corrector problem.  However, from an engineering point of view, modifying the underlying partial differential equation for more general problems to be homogenized, e.g.\ elasticity or inelasticity, can only be a last resort, even if it improves the convergence rate. Apart from physical reasons, such an approach requires modifying existing computational homogenization codes at their core.}\\
Recently, Clozeau et al.~\cite{CloJoOtXu_2020} {worked out the details of an attractive} alternative to {modifying the} corrector problem.
{Instead, they modify the ensemble by periodizing it on cells of finite size, thus recovering the more favorable $\review{1/L^{d}}$ convergence rate of the systematic error (without modifying the corrector equation). This is in line with the existing bounds of Gloria et al.~\cite{GloriaNeukammOtto2015} and Egloffe et al.~\cite[Sec. 3.2.3]{Egloffe2015}, which concern i.i.d.\ random conductivities on a discrete lattice, and with Khoromskaia et al.~\cite{Khoromskiis2019}, who investigated an ensemble based on a discrete Poisson point process numerically.
However, periodizing the aforementioned ensembles is trivial; on the contrary, Clozeau et al.~\cite{CloJoOtXu_2020} deal with a less academic and more realistic class of ensembles, namely coefficient fields generated by applying a nonlinear transformation to a Gaussian field with (essentially) integrable correlation function.
Moreover, Clozeau et al.~\cite{CloJoOtXu_2020} do not only provide an estimate of the systematic error but actually derive the leading-order term in $L^{-d}$, which is generically non-vanishing.}
{For applications in} engineering, {the} result~\cite{CloJoOtXu_2020} suggests that working with {periodized ensembles} leads to a favorable {systematic error} compared to the {snapshot} approach. This is very much in line with observations in engineering. For instance, considering Poisson-Voronoi microstructures, Kanit et al.~\cite[Sec. 7]{KanitRVE} observe that "the bias {[systematic error]} introduced by the periodic boundary conditions is found to be much smaller than for the other boundary conditions".

\subsection{Contributions}

We carry out numerical experiments to quantify the systematic and the random error for composite microstructures. In contrast to previous works~{\cite{Khoromskiis2019, Khoromskiis2020}}, we use microstructures of industrial relevance and complexity, utilizing real physical parameters{, investigating spherical and cylindrical fillers}. {While the ensembles considered here do not fall into the class considered in the theory - they are neither finite range nor a suitable functional inequality is known, see, however, Duerinckx \& Gloria~\cite{WeightedFunctionalInequalities,WeightedFunctionalInequalities2} - we expect that they belong to the same universality class.} Thus, the composites under consideration serve as benchmarks for the established quantitative homogenization theory, and are expected to stimulate further theoretical research.\\
{We} are interested in the consequences of improper treatment of particles intersecting the boundary of computational volume elements, leading to the boundary-layer error, and compare these results to properly periodized {ensembles}.\\
From a practical engineering perspective, these results shed light on modern up-scaling techniques based on digital volume images, for instance obtained by micro-computed tomography. Indeed, the latter images naturally correspond to {snapshots of ensembles with} "improperly" treated particles at the boundaries. The effort of extracting such a specimen precludes computing the apparent properties corresponding to this cell size, as the necessary number of samples, which is typically on the order of $10^5$ for engineering accuracy~\cite{Khoromskiis2019, Khoromskiis2020}, is both too expensive and excessively arduous. For this purpose, we prepare a similar protocol under "laboratory conditions". More precisely, we consider the simplified scenario of thermal conductivity, but rely upon modern computational homogenization techniques based on the fast Fourier transform (FFT), pioneered by Moulinec \& Suquet~\cite{MoulinecSuquet1994,MoulinecSuquet1998}, and validated methods for generating synthetic microstructure-generation algorithms for spherical and cylindrical fillers~\cite{WilliamsPhilipse,SAM} to quantify the effect of {taking snapshots of} random media {on} finite-sized cells.\\
In {Section} \ref{sec:theory}, we provide a streamlined presentation of quantitative stochastic homogenization suitable for readers with an engineering background. After exposing the microstructure-generation tools in {Section} \ref{sec:microstructure_gen}, the computational results are presented in {Section} \ref{sec:results}, comprising over $250\,000$ simulation runs with up to more than one billion degrees of freedom. An extensive discussion of the non-standard decay of the {random} error is provided in {Section} \ref{sec:discussion}.

\section{Theoretical background}
\label{sec:theory}

	\begin{figure}
		\begin{subfigure}{.49\textwidth}
			\includegraphics[width=\textwidth]{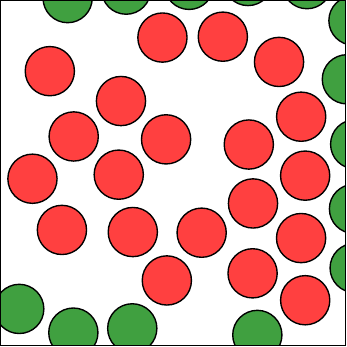}
			\caption{{Taking a snapshot} introduces cut particles ({green}) in addition to original particles ({red})}
			\label{fig:periodization_nonperiodic}
		\end{subfigure}
		\begin{subfigure}{.49\textwidth}
			\includegraphics[width=\textwidth]{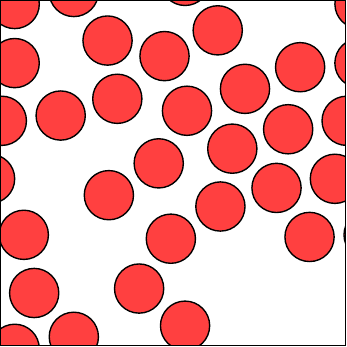}
			\caption{For a properly periodized ensemble, all inclusions belong to the same species}
			\label{fig:periodization_periodic}
		\end{subfigure}
		\caption{Illustrating {snapshot} coefficients $A^{{\textrm{sn}}}_L$ vs. periodized coefficients $A^\text{per}_L$}
		\label{fig:periodization}	
	\end{figure}
We are concerned with a stationary and ergodic random field of thermal conductivity tensors $A$ in $d$ spatial dimensions, i.e., every realization is a field on $\R^d$ with values in symmetric positive definite $d\times d$-matrices. In stochastic homogenization~\cite{PapanicolaouVaradhan, Kozlov}, the effective conductivity tensor $\bar{A}$ is sought, and may be determined for any prescribed (negative) temperature gradient $\bar{\xi} \in \R^d$ via the expectation of the heat flux
\begin{equation}\label{eq:effective_conductivity_stochastic}
	\bar{A}\bar{\xi} = \mean{A \xi},
\end{equation}
where $\xi$ denotes the corresponding \emph{local} (negative) temperature gradient, i.e.,
\begin{equation}\label{eq:xi_compatible_stochastic}
	\xi = \bar{\xi} + \Gamma \xi
\end{equation}
holds in terms of the Helmholtz projector $\Gamma$ onto {stationary} gradient fields, and the associated heat flux $q = A\xi$ is divergence-free, i.e., {the equation}
\begin{equation}\label{eq:div_q_stochastic}
	\Gamma q = 0
\end{equation}
is satisfied. Due to the ergodicity assumption, the effective conductivity tensor $\bar{A}$ may also be computed in terms of a single realization involving an infinite{-}volume limit, see Chapter 7 in Zhikov et al.~\cite{zhikov1994homogenization} for details.\\
For computational purposes it is {necessary} to work on finite domains. {For a fixed cube $Q_L = \left[-\tfrac{L}{2},\tfrac{L}{2}\right]^d$, assume that we are given an ensemble $A_L$ that generates coefficient fields on the cell $Q_L$. The latter can be obtained {as snapshots of} the random conductivity tensor field $A$, which is given on the whole space, {on} the cube $Q_L$. In this case, we will write $A^{{\textrm{sn}}}_L$ for the corresponding ensemble. An alternative approach proceeds by \textit{periodizing} the ensemble. In this case, we will denote the ensemble by $A^\text{per}_L$. {Taking snapshots of} an ensemble is straightforward because it can be performed on given realizations, whereas periodization is a more subtle process (to be discussed below) {and} requires working on the level of the ensemble itself. {However, periodization is quite} natural when generating synthetic microstructures.\\
%
%
Let us illustrate what we mean by periodizing the {ensemble}. For i.i.d.\ discrete coefficient fields~{\cite{GloriaNeukammOtto2015} and the Poissonian ensemble~\cite{GloriaOtto2015QuantitativeEstimates}}, this procedure is trivial. For Gaussian fields, periodization amounts to periodizing the correlation function, see Clozeau et al.~\cite{CloJoOtXu_2020}.
\begin{figure}
	\begin{center}
		\includegraphics[width=.4\textwidth]{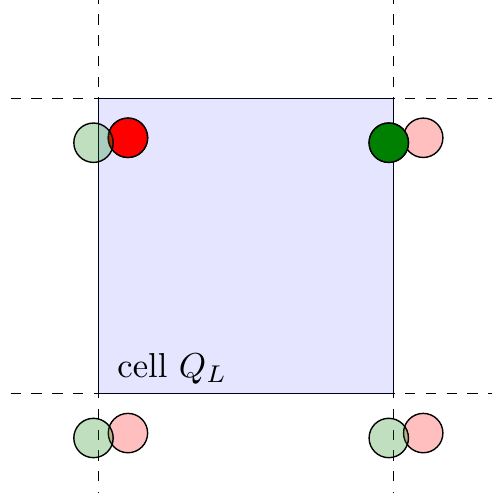}
		\caption{{Periodizing the distance for a two-dimensional cell $Q_L$ containing two circles (red and green) of equal radius and their shaded periodically replicated images}}
		\label{fig:periodic_distance}
	\end{center}
\end{figure}
	For the article at hand, we consider a particle-reinforced composite with a single species of particles, for instance spherical particles, and specific {macroscopic} statistics. When {taking a snapshot of} the coefficient field {on} a box $Q_L$ (thus building $A^{{\textrm{sn}}}_L$), particles intersecting the boundary of $Q_L$ are cut, giving rise to another species of particles featuring its own statistics, see Fig.~\ref{fig:periodization_nonperiodic}. In particular, {taking snapshots of} the particles does not retain the stationarity property of the {whole}-space ensemble.\\
	To understand how periodization proceeds let us look at the way an ensemble of particle-filled composites is built. In real life, the fillers do not overlap (unless this is part of the production process, for instance in sintering). Thus, when generating an ensemble, an overlap check needs to be made. For the periodized ensemble, the particles are placed in a fixed cell $Q_L$, and the overlap checks are made not only between the original particles, but also with their periodic translations, see Fig.~\ref{fig:periodic_distance} for an illustration. Following such a protocol with statistical properties properly matching the {whole}-space counterpart, see Fig.~\ref{fig:periodization_periodic}, gives rise to the \textit{periodized} ensemble and the associated coefficient field $A_L^\text{per}$, which is stationary.
	As will be shown below, the cut particles of $A^{{\textrm{sn}}}_L$ are considered responsible for deteriorating the quality of the computed homogenized coefficient with respect to the periodized coefficient field $A_L^\text{per}$ (see eq. \eqref{eq:systematic_error_convergence} versus eq. \eqref{eq:systematic_error_convergence2}).
}
\\
In {either} case, we denote by $\bar{A}_L$ the apparent conductivity of $A_L$ on $Q_L$ endowed with periodic boundary conditions, i.e., in analogy to the stochastic {setting} \eqref{eq:effective_conductivity_stochastic}, for any $\bar{\xi} \in \R^d$, 
\begin{equation}\label{eq:effective_conductivity_periodic}
	\bar{A}_L\bar{\xi} = \dashint_{Q_L} A_L \xi_L \, dx,
\end{equation}
where $\xi_L \in L^2(Q_L;\R^d)$ solves the equations
\begin{equation}\label{eq:xi_compatible_div_q_periodic}
	\xi_L = \bar{\xi} + \Gamma_L \xi_L \quad \text{and} \quad \Gamma_L A_L \xi_L  = 0
\end{equation}
in terms of the Helmholtz projector $\Gamma_L$ onto gradient fields of periodic functions on the cube $Q_L$. Notice that the apparent conductivity tensor $\bar{A}_L$ is still a random variable. Also, its {expectation} $\mean{\bar{A}_L}$ will be different from the effective conductivity $\bar{A}$, in general, which is well-known in the engineering community, see Huet~\cite{Huet1990} and Sab~\cite{Sab1992} for early accounts. However, it can be shown, {either in case $A_L=A^{{\textrm{sn}}}_L$~\cite{BourgeatPiatnitski,Owhadi2003} or in case $A_L=A_L^\text{per}$, see Gloria et al.~\cite{GloriaNeukammOtto2015} and Clozeau et al.~\cite{CloJoOtXu_2020} for specific frameworks}, that 
\begin{equation}\label{eq:Atr_convergence_qualitative}
	\bar{A}_L \rightarrow \bar{A} \quad \text{as} \quad L \rightarrow \infty
\end{equation}
with probability one, i.e., the apparent conductivity tensors $\bar{A}_L$ approximate the effective conductivity tensor $\bar{A}$ for increasingly large cubes $Q_L$. In the engineering community, a sufficiently large cell $Q_L$ is called a \emph{representative volume element}~\cite{HillRVE, KanitRVE}. {Note} that boundary conditions different from periodic could be used as well for defining the truncated corrector problem \eqref{eq:xi_compatible_div_q_periodic}, e.g., Dirichlet or Neumann boundary conditions, but we chose periodic boundary conditions as they tend to be more accurate for fixed cell size, see, for instance, Kanit et al.~\cite{KanitRVE}.\\
The convergence statement \eqref{eq:Atr_convergence_qualitative} is purely qualitative and does not permit {to analyze the effect of the size of the} volume elements. Indeed, in practice a suitable trade-off between the increased accuracy that accompanies larger computational cells and the limited computational resources needs to be identified, {for instance based on} statistical estimation {techniques}~\cite{KanitRVE}. For this purpose, \emph{quantitative convergence statements} become useful. On average, the error may be decomposed in the form
\begin{equation}\label{eq:homogenization_error_decomposition}
	{\mean{ \left\| \bar{A}_L - \bar{A}\right\|^2 } =  \mean{ \left\| \bar{A}_L - \mean{\bar{A}_L}\right\|^2 } + \left\| \mean{ \bar{A}_L } - \bar{A} \right\|^2}
\end{equation}
{in terms of the Frobenius norm}, i.e., the right-hand side consists of a random error, which measures the standard deviation of the {snapshot} ensemble, and a systematic error which quantifies the defect induced by working on a cell of finite size. From a physical point of view, the random error quantifies the lack of statistical representativity of the {realization in the} finite-sized cell, whereas the systematic error accounts for artificial long-range {correlations} induced by {periodization for $A_L^\text{per}$ and the boundary-layer error in case of {taking snapshots} $A^{{\textrm{sn}}}_L$}.\\
By using a Monte-Carlo sampling-strategy on cells of fixed size, the random contribution may be decreased - in contrast to the systematic error. Indeed, suppose that $N$ {independent} coefficient fields $\left( A_{L,i} \right)$ are sampled. Then, the empirical {expectation} of the corresponding effective conductivities satisfies
\[
	{
	\mean{\left\| \frac{1}{N} \sum_{i=1}^N \bar{A}_{L,i} - \bar{A}\right\|^2} = \frac{1}{N}\,\mean{ \left\| \bar{A}_L - \mean{\bar{A}_L}\right\|^2 } + \left\| \mean{ \bar{A}_L} - \bar{A} \right\|^2.}
\]
The decomposition \eqref{eq:homogenization_error_decomposition} is implicit in the statistical approach to estimating the minimum size of a representative volume element pioneered by Kanit et al.~\cite{KanitRVE}. They assume the systematic error, which they call \emph{bias}, to be negligible compared to the {random} error, which they refer to as \emph{dispersion}.\\
In general, due to the central limit theorem (CLT), we cannot expect a decay of the random error better than
\begin{equation}\label{eq:random_error_CLT}
	\sqrt{ \mean{ \left\| \bar{A}_L - \mean{\bar{A}_L}\right\|^2 } } \lesssim L^{-\frac{d}{2}}.
\end{equation}
{This scaling was confirmed by Kanit et al.~\cite{KanitRVE} in computational experiments for synthetic Poisson-Voronoi microstructures and thermal conductivity. In case of elasticity, a slightly inferior scaling was observed.}\\
{On the theoretical side, notice that the decay of the error \eqref{eq:homogenization_error_decomposition} may be arbitrarily slow for general stationary and just qualitatively ergodic coefficient fields. 
{Specific quantified ergodicity assumptions permit drawing stronger conclusions. For instance, ensembles with finite range were considered by Armstrong \& Smart~\cite{ArmstrongSmart}, who provided quantitive stochastic homogenization results for convex integral functionals of quadratic growth. Armstrong \& Mourrat~\cite{ArmstrongMourrat} relaxed the finite-range assumption and provided results for a mixing condition with algebraic decorrelation rate, previously considered by Yurinskii~\cite{Yurinskii1986}.\\ 
As an alternative to the approaches just mentioned, ergodicity may be quantified in terms of suitable functional inequalities in probability~\cite{NaddafSpencer}, which compensate the lack of "natural" Poincaré inequality on the probability space.} Under a{n $L^2$} spectral gap assumption, either if $A_L$ is obtained {as a snapshot}~\cite[eq.\ (3.3)]{Egloffe2015} or by periodization~\cite[Th.\ 1]{QTensor}~\cite[Th.\ 2]{GloriaNeukammOtto2015}, the CLT estimate \eqref{eq:random_error_CLT} is expected to hold. For the latter case, Duerinckx et al.~\cite[Th.\ 1]{QTensor} actually identify the first order term in the expansion of $\bar{A}^\text{per}_L - \mean{\bar{A}^\text{per}_L}$. As a consequence, the CLT scaling \eqref{eq:random_error_CLT} of the random error is indeed optimal.\\
However, these quantifications of ergodicity only hold under restrictive assumptions on the underlying random coefficient field and may not meet the requirements of practical models of interest to the applied sciences~\cite{Torquato}. {Thus, spectral gap assumptions allowing for thicker stochastic tails were considered, i.e., in terms of an $L^p$ spectral gap assumption~\cite{FischerOtto2017} and via a weighted logarithmic Sobolev inequality~\cite[Def. 1]{GloriaNeukammOttoLSI2019}.} {Further weighted} functional inequalities {satisfied by ensembles of relevance to materials science were studied by Duerinckx \& Gloria}~\cite{WeightedFunctionalInequalities,WeightedFunctionalInequalities2}.\\
In contrast to the random error, the convergence rate of the systematic error is sensitive to the way the finite-cell ensemble $A_L$ {is build}.
In general, a boundary-layer error {(as introduced earlier)} with dimension-independent scaling
\begin{equation}\label{eq:systematic_error_convergence}
\left\| \mean{ \bar{A}^{{\textrm{sn}}}_L } - \bar{A} \right\|  \lesssim L^{-1}
\end{equation}
is expected, see Egloffe et al \cite[Eq.\ (3.4)]{Egloffe2015}, and proved {to be} optimal in Clozeau et al.~\cite{CloJoOtXu_2020} for a specific class of coefficient fields.
If the coefficient fields  $A^{{\textrm{sn}}}_L$ are considered as given, techniques based on modifying the corrector equation \eqref{eq:xi_compatible_div_q_periodic} were proposed~\cite{FancyStuff1,FancyStuff2} to screen the influence of the improperly treated boundary of the considered cell and to restore a better convergence rate.
In contrast, using the periodized coefficient field $A_L=A_L^\text{per}$ may {attenuate} the systematic error, so that {screening} techniques are not necessary.\\
The latter statement could be made rigorous for specific classes of ensembles {that allow for a natural periodization}.
For i.i.d.\ discretely random conductivities, Gloria et al.~\cite[Th.\ 2]{GloriaNeukammOtto2015} established the estimate
\begin{equation}\label{eq:systematic_error_convergence2}
\left\| \mean{ \bar{A}_L^\text{per} } - \bar{A} \right\| \lesssim L^{-d}
\end{equation}
for $d>2$, where there is an additional factor of $\ln^d(L)$ in the r.h.s.  Such a result {is known to hold} for the case of overlapping Poisson random inclusions. For the case where the ensemble is generated from a Gaussian field with integrable covariance function, Clozeau et al.~\cite{CloJoOtXu_2020} showed estimate \eqref{eq:systematic_error_convergence2}, and proved this scaling to be optimal in this setting by identifying the first order term of the asymptotic expansion of $\mean{ \bar{A}_L^{{\text{per}}}} - \bar{A}$.}

\section{Computational tools}
\label{sec:microstructure_gen}

In this {Section} we provide details on the microstructure-generation methods used as the basis for the computational experiments conducted in {Section} \ref{sec:results}.

\subsection{The mechanical contraction method}
\label{sec:microstructure_gen_spheres}

In this {Section} we describe the mechanical-contraction method (MCM) of Williams-Philipse~\cite{WilliamsPhilipse} used for generating circle packings (in 2D) and sphere packings (in 3D). For composites with spherical fillers, the simple random sequential adsorption algorithm of Feder~\cite{Feder1980} is unable to reach industrial-scale volume fractions in reasonable time. Indeed, the jamming limit of mono-sized spheres is approximately $38\%$~\cite{RSAJam}, which is insufficient for the computational experiments of {Section} \ref{sec:results_setup} below, taking into account the isolation distance.\\
For this purpose, collective rearrangement algorithms are necessary, and several methods are available~\cite{LubachevskyStillinger, TorquatoJiao}. We shall describe the mechanical-contraction method, as there is a modification which generalizes to cylindrical fillers.\\
Suppose that a radius $r$ and a cubic cell $Q_L$ are given, together with an initial distribution of midpoints $x_1,\ldots,x_N$, each contained in $Q_L$. Suppose that $r$, $L$ and $N$ are chosen in such a way that, if all spheres centered at the midpoints are non-overlapping, the target volume fraction $\phi$ would be reached. If the spheres are overlapping, the MCM applies an overlap-removal technique in order to move the midpoints $x_1,\ldots,x_N$ in such a way that the spheres do not overlap anymore. For this purpose, an overlap energy
\begin{equation}\label{eq:mechanical_contraction_overlap_energy}
	W: Q_L^N \rightarrow \R, \quad (x_1,\ldots,x_N) \mapsto \frac{1}{2} \sum_{1\le i < j \le N} \delta(x_i,x_j)^2,
\end{equation}
is defined in terms of an overlap indicator
\begin{equation}\label{eq:mechanical_contraction_defn_delta}
  \delta(x_i,x_j) = \langle 2r - \textrm{dist}(x_i,x_j) \rangle_+,
\end{equation}
where $\textrm{dist}(x_i,x_j)$ denotes the $Q_L$-periodic distance of $x_i$ and $x_j$ and 
\[
  \langle z \rangle_+ = \max(0,z)
\]
is the Macauley bracket.
\begin{figure}
	\begin{subfigure}{.32\textwidth}
		\includegraphics[width = \textwidth]{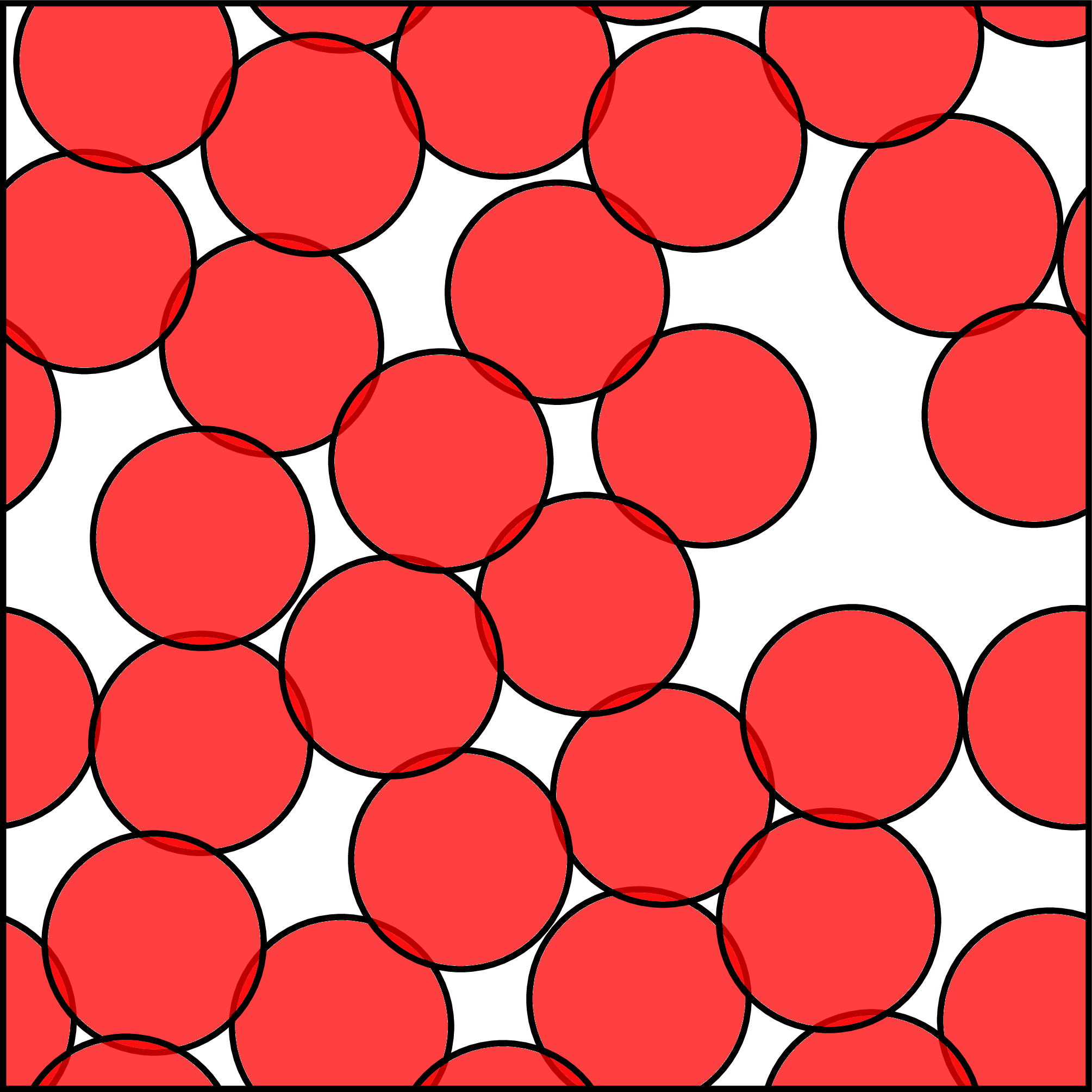}
		\caption{Initial configuration}
		\label{fig:MCM_1}
	\end{subfigure}
	\begin{subfigure}{.32\textwidth}
		\includegraphics[width = \textwidth]{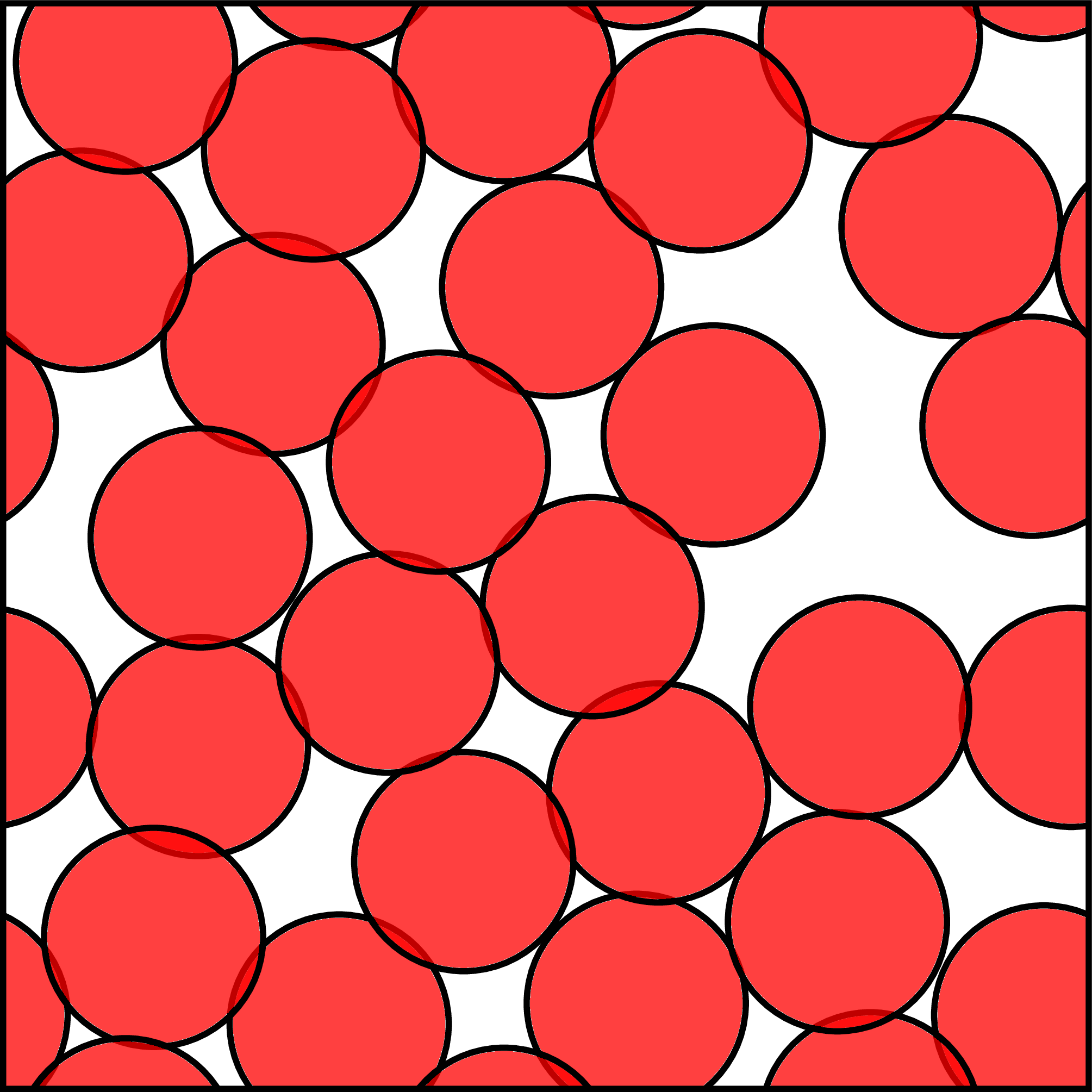}
		\caption{After 1 step}
		\label{fig:MCM_2}
	\end{subfigure}
	\begin{subfigure}{.32\textwidth}
		\includegraphics[width = \textwidth]{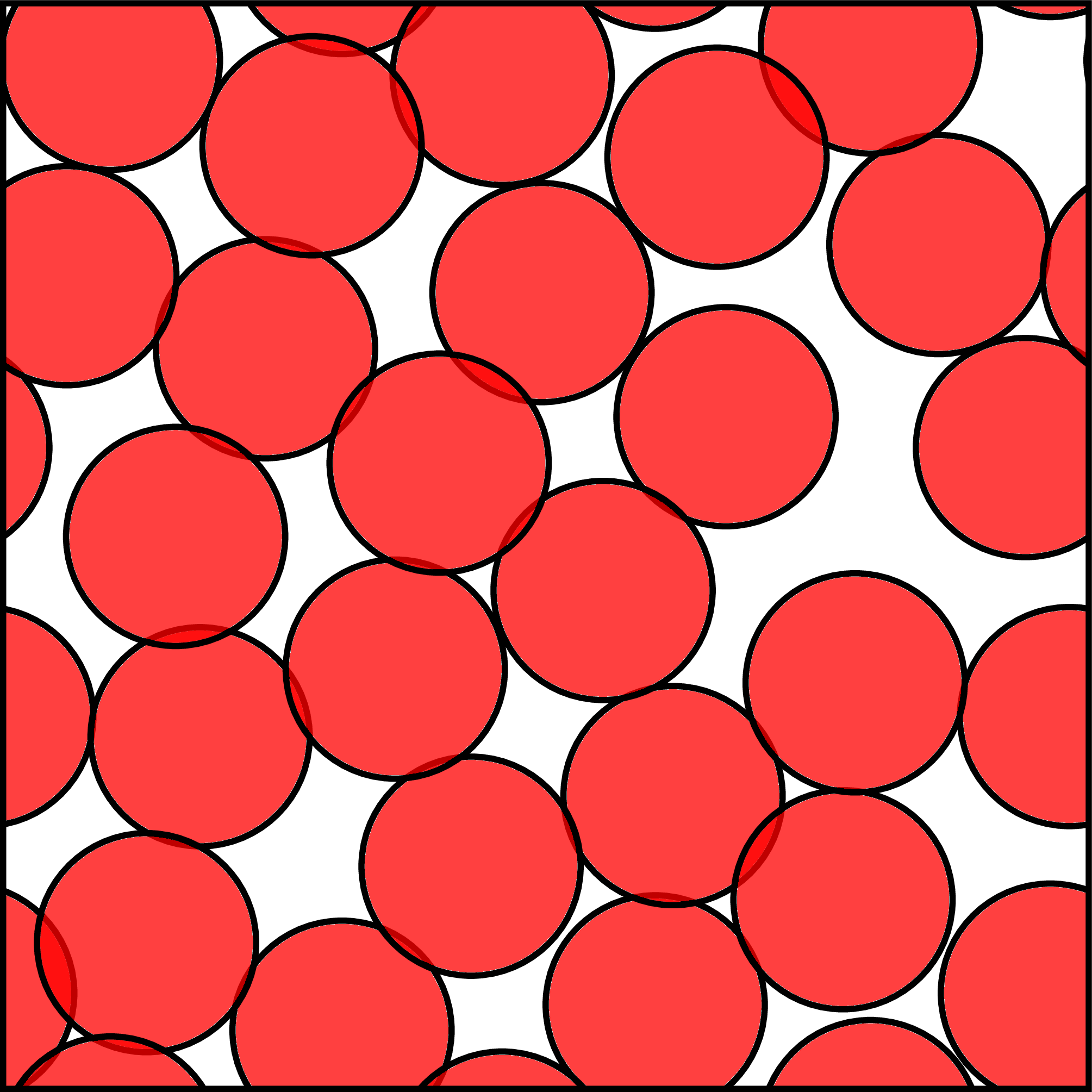}
		\caption{After {10} steps}
		\label{fig:MCM_3}
	\end{subfigure}\\
	\begin{subfigure}{.32\textwidth}
		\includegraphics[width = \textwidth]{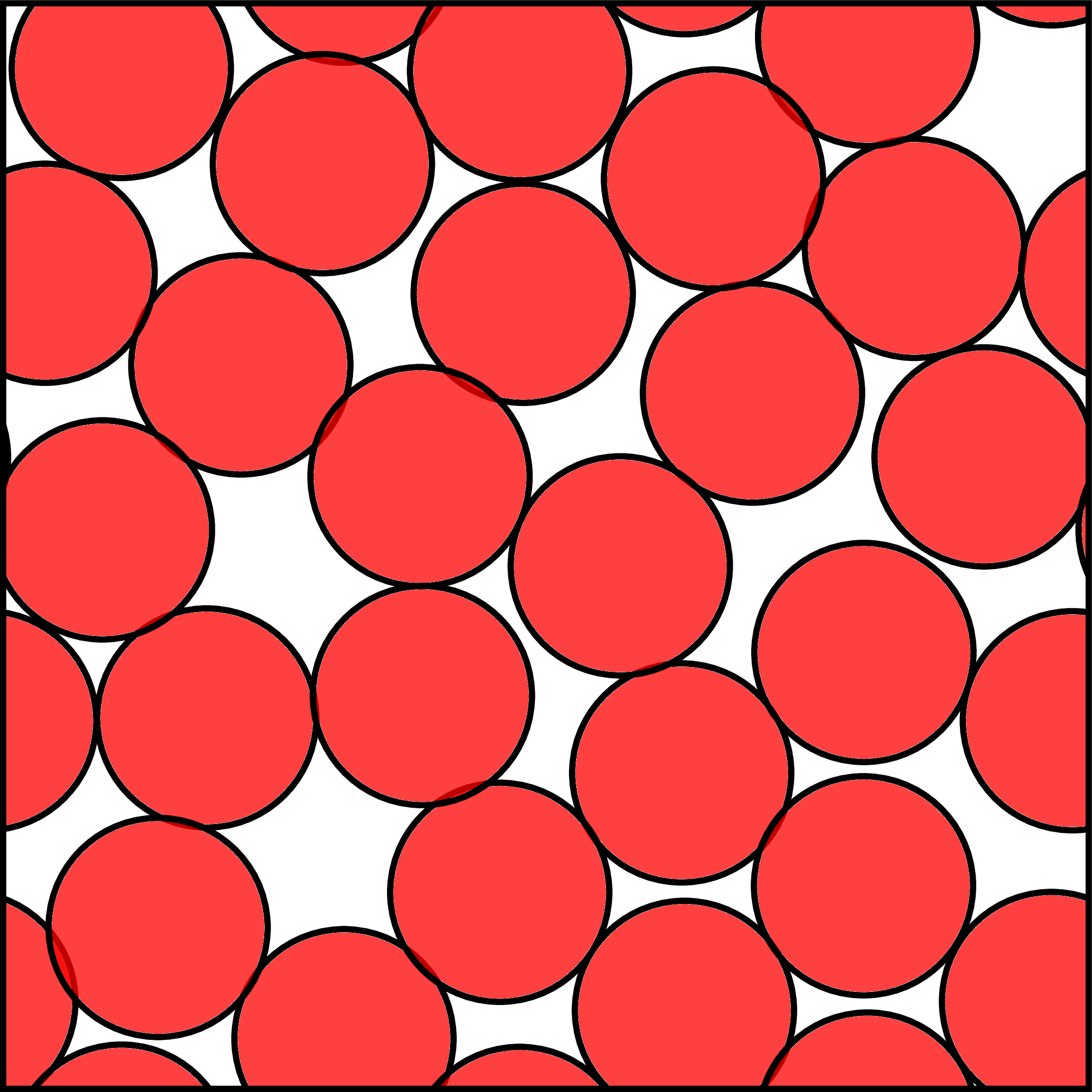}
		\caption{After {50} steps}
		\label{fig:MCM_4}
	\end{subfigure}
	\begin{subfigure}{.32\textwidth}
		\includegraphics[width = \textwidth]{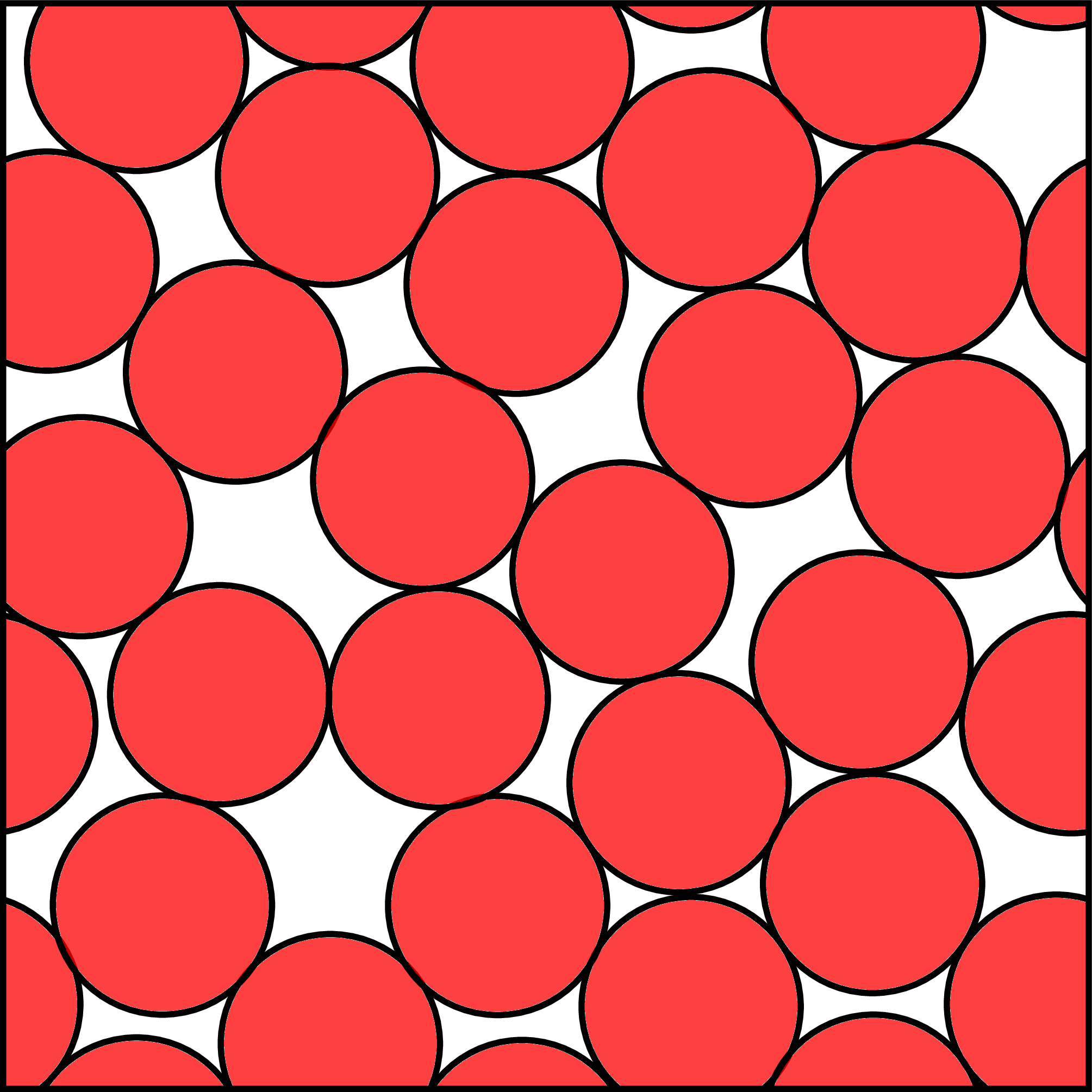}
		\caption{After {100} steps}
		\label{fig:MCM_5}
	\end{subfigure}
	\begin{subfigure}{.32\textwidth}
		\includegraphics[width = \textwidth]{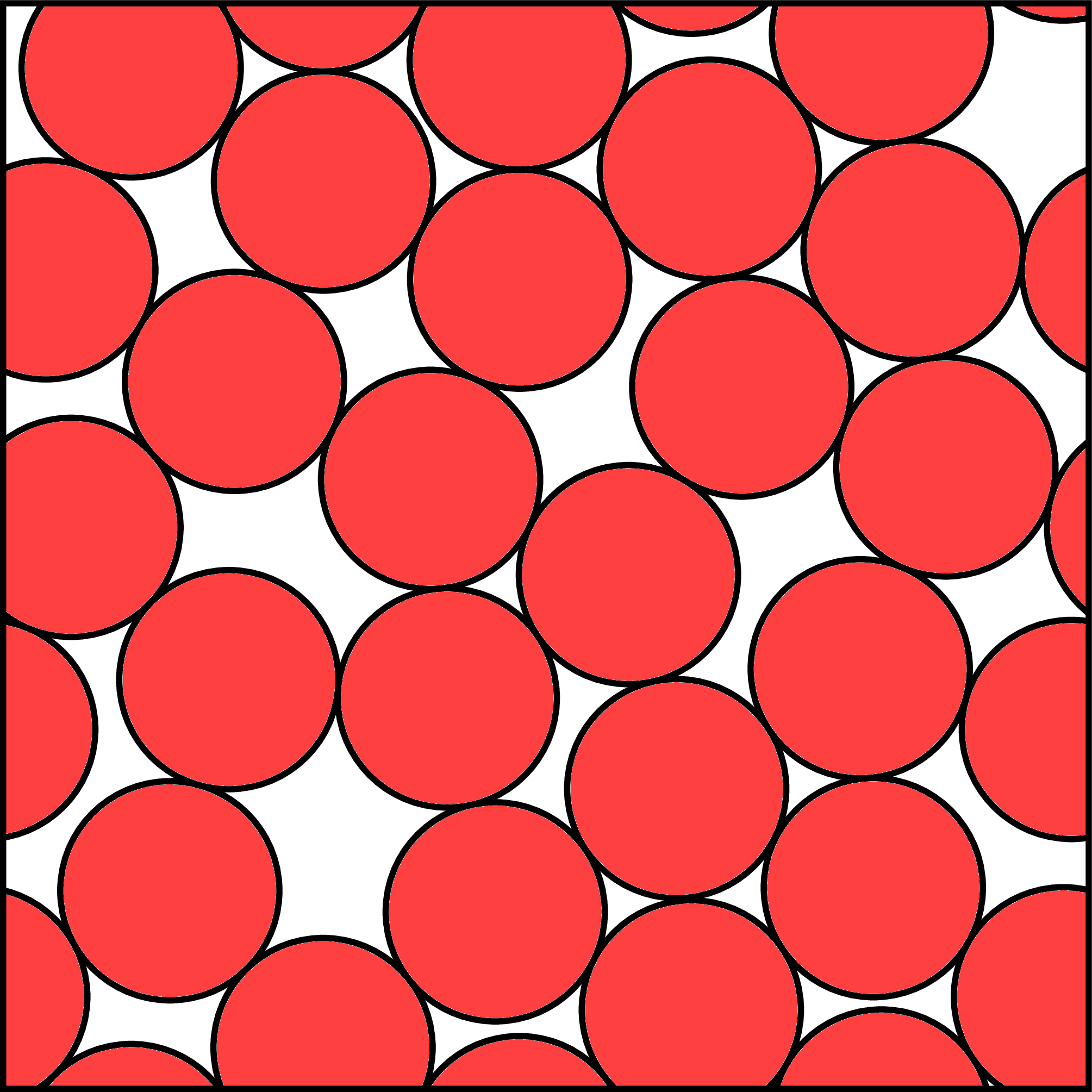}
		\caption{Final configuration}
		\label{fig:MCM_final}
	\end{subfigure}\\
	\caption{Illustration of the overlap-removal technique central to the mechanical contraction method~\cite{WilliamsPhilipse}, applied to {$25$} circles {and an area fraction of $80\%$}}
	\label{fig:MCM}
\end{figure}
Two (open) spheres with radius $r$ centered at $x_i$ and $x_j$, respectively, do not overlap if and only if $\delta(x_i,x_j) = 0$. Thus, the $N$ spheres centered at $x_1,\ldots,x_N$ are in a non-overlapping configuration precisely if $W(x_1,\ldots,x_N) = 0$. As the overlap energy $W$ \eqref{eq:mechanical_contraction_overlap_energy} is continuously differentiable, a gradient-descent method may be used for finding a global minimizer of $W$, see Williams-Philipse~\cite{WilliamsPhilipse} for details. The approach is illustrated in Fig.~\ref{fig:MCM}, where {$25$} disks were placed in an initially random configuration at ${8}0\%$ volume fraction. During the iterations, the centers of overlapping disks are moved, whereas disks without overlap remain fixed. For this example, the overlap-removal algorithm converged to the desired precision after ${288}$ iterations.\\
In the MCM algorithm, an increasing sequence of volume fractions $\phi_1,\phi_2,\ldots, \phi_K$ is defined. For the initial volume fraction $\phi_1$ (which is typically rather low), the initial centers of the spheres are drawn from a uniform distribution on the corresponding cube. Then, the overlap-removal technique is applied. Once a non-overlapping configuration is reached for $\phi_k$, both the cell and the sphere centers are rescaled in such a way that the non-overlapping configuration would have a volume fraction $\phi_{k+1}$. Then, the overlap-removal technique is applied. The process is repeated until the desired volume fraction $\phi_K$ is reached. The repeated shrinking is responsible for the name of the algorithm: mechanical contraction method. Several remarks are in order.\\
\begin{enumerate}
	\item The complexity of the MCM algorithm is hidden in computing the pairwise distances \eqref{eq:mechanical_contraction_defn_delta}. A naive implementation requires on the order of $N^2$ computations, which can be excessive for large $N$. We rely upon cell-linked lists~\cite{CellLists} to reduce the complexity to $O(N)$.
	\item In practice, an isolation distance between the spheres is useful to avoid singularities in the solution field. This is realized by working with a slightly larger radius $\tilde{r}>r$ when computing the overlap indicator \eqref{eq:mechanical_contraction_defn_delta}.
\end{enumerate}

\subsection{Sequential addition and migration}
\label{sec:microstructure_gen_fibers}

The mechanical contraction method of Williams-Philipse~\cite{WilliamsPhilipse} also applies to spherocylinders, i.e., cylinders with spherical caps attached, provided the overlap indicator \eqref{eq:mechanical_contraction_defn_delta} is modified. A microstructure generated in this way may be used as a model of a fiber-reinforced composite by neglecting the spherical caps of the spherocylinders.\\
In its original form, however, the method can only generate moderate volume fractions without altering the overall fiber orientation. For this reason, Schneider~\cite{SAM} modified the overlap energy $W$ \eqref{eq:mechanical_contraction_overlap_energy} by a penalty term accounting for the orientation. Also, in order to reach higher volume fractions, instead of successively shrinking the cell, fibers are added incrementally to a cell of fixed size. The resulting algorithm is called sequential addition and migration (SAM), see Schneider~\cite{SAM} for details.
%

\section{Computational results}
\label{sec:results}

\subsection{Setup}
\label{sec:results_setup}

We are concerned with microstructures of particle{-}reinforced composites. We used the isotropic thermal conductivity of polypropylene ($0.2 \text{W/(m$\cdot$K)}$) for the matrix and of E-glass ($1.2 \text{W/(m$\cdot$K)}$) for the fillers, see Dorn-Schneider~\cite{DornSchneider2019}.\\
The microstructures were discretized on a regular pixel/voxel grid, segmented according to which phase the midpoint of the pixel/voxel belongs to. The static thermal equilibrium problem \eqref{eq:xi_compatible_div_q_periodic} was discretized by the Moulinec-Suquet discretization~\cite{MoulinecSuquet1994,MoulinecSuquet1998}, an underintegrated Fourier-Galerkin discretization~\cite{VondrejcFFTGalerkin}, see Brisard-Dormieux~\cite{BrisardDormieux2012} and Schneider~\cite{MSConvergence2015} for convergence proofs. For resolving the discrete system, we used the Eyre-Milton solver~\cite{EyreMilton1999}, a Douglas-Rachford type {solution method}~\cite{EyreMilton2019}, with a tolerance of $10^{-6}$, implemented in Python with Cython extensions, see Schneider~\cite{BB2019} for more details.\\
For the circular and spherical inclusions, a desktop computer with a 6-core Intel i7 CPU and 32GB RAM was utilized. A dual-socket workstation with $48$ cores and $1$ TB RAM was used for the short-fiber microstructures.

\subsection{Spherical inclusions}
\label{sec:results_spheres}

\begin{figure}
	\begin{subfigure}[t]{.066\textwidth}
		\includegraphics[width=\textwidth, trim = 540 100 440 250, clip]{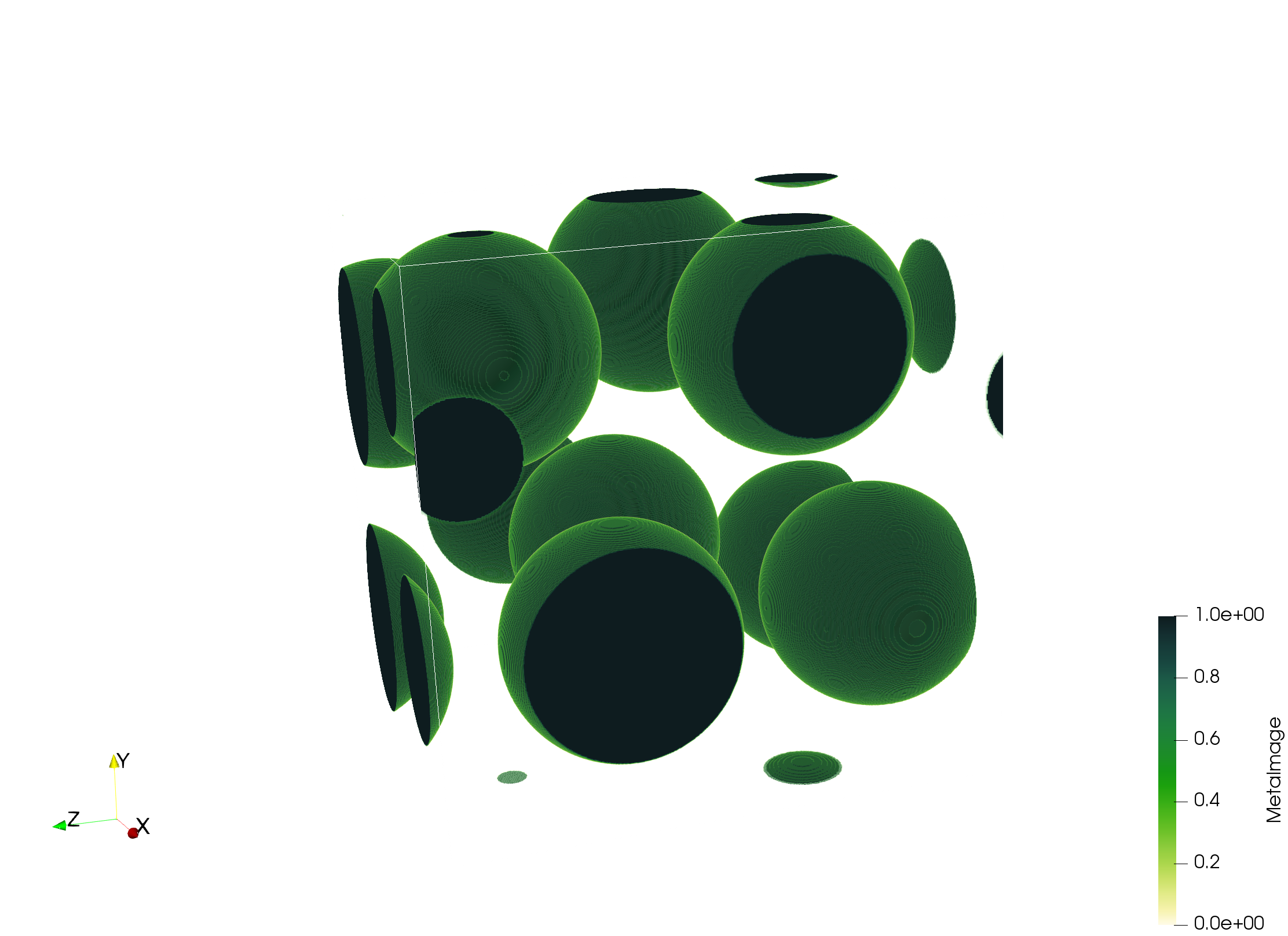}
	\end{subfigure}		
	\begin{subfigure}[t]{.132\textwidth}
		\includegraphics[width=\textwidth, trim = 540 100 440 250, clip]{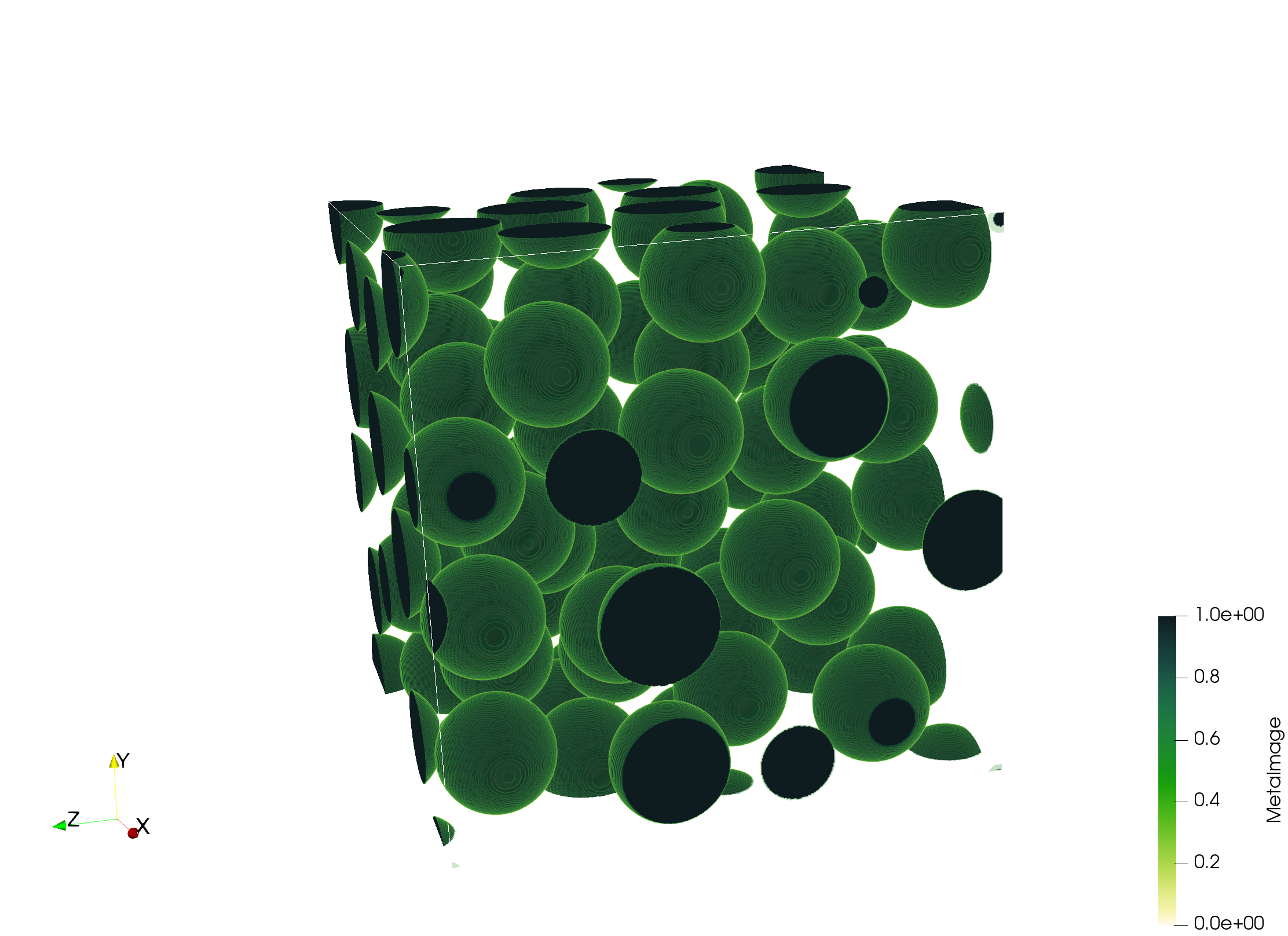}
	\end{subfigure}	
	\begin{subfigure}[t]{.264\textwidth}
		\includegraphics[width=\textwidth, trim = 540 100 440 250, clip]{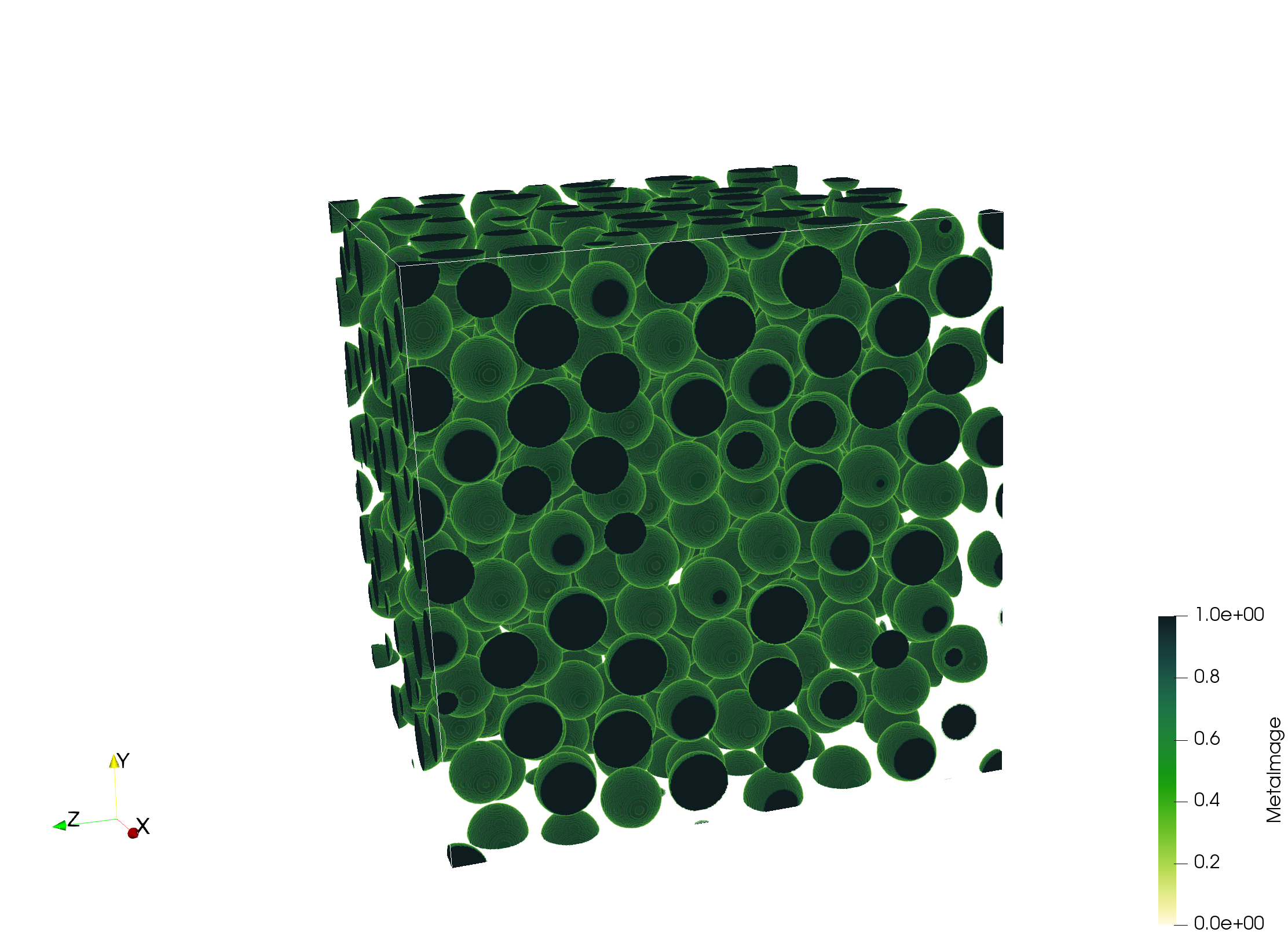}
	\end{subfigure}	
	\begin{subfigure}[t]{.528\textwidth}
		\includegraphics[width=\textwidth, trim = 540 100 440 250, clip]{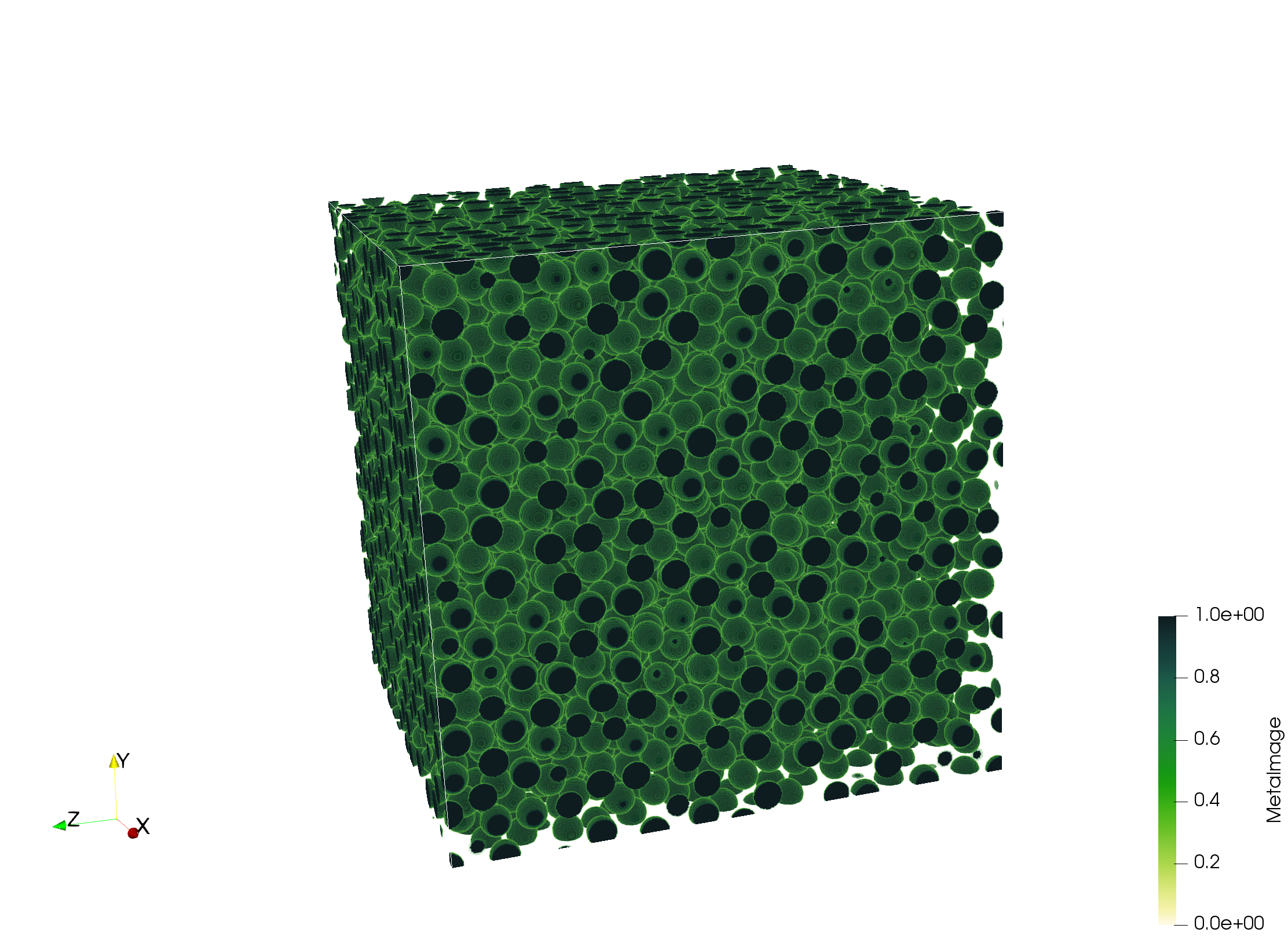}
	\end{subfigure}				
	\caption{3D comparison of periodically generated cells and increasing number of inclusions, with $2^3 = 8$, $4^3 = 64$, $8^3 = 512$ and $16^3 = $\numprint{4096} spheres (from left to right)}
	\label{fig:spheres_illustration_3D}
\end{figure}
We investigate mono-{disperse} spherical fillers as our first example. For generating microstructures, we follow two protocols. The first strategy, which we call \emph{periodized}, generates $K^3$ ($K=2,4,8,16$) non-overlapping spheres by the mechanical contraction method, see {Section} \ref{sec:microstructure_gen_spheres}. For the second, {\emph{snapshot}} strategy, $(2K)^3$ non-overlapping spheres are packed as before, but we cut out $1/8$th of the structure. Thus, we arrive at a non-periodic structure of identical size as for the previous protocol.\\
For both approaches, we fill to $30\%$ by the mechanical contraction method of Williams and Philippse~\cite{WilliamsPhilipse} in $10\%$ steps, and we use an isolation distance of $20\%$ of the sphere's radius.\\
A unit cell containing $K^3$ spheres (on average) is discretized by $(16 K)^3$ voxels, i.e., the microstructures for $K=16$ are discretized by $256^3=$\numprint{16777216} elements and contain \numprint{4096} spheres. Following Khoromskaia et al.~\cite{Khoromskiis2019}, {Section} 5, we generated \numprint{10000} microstructures for $K=2,4,8,16$ and the two protocols. The most time-consuming step here was generating the {snapshot} elements with $K=16$, because \numprint{10000} volume elements{, each} containing \numprint{32768} {non-overlapping} spheres{,} needed to be generated quickly.\\
For each of the \numprint{10000} generated microstructures, we computed the $11$-component of the effective thermal conductivity. To keep notation simple, we introduce
\begin{equation}\label{eq:a_bar}
	\bar{a} = \bar{A}_{11}
\end{equation}
as our quantity of interest. In contrast to the volume fraction, the ground truth for $\bar{a}$ is not known exactly. For that reason, we need to resort to computations on a suitable large scale. More precisely, we generated ten large {microstructures with the periodization strategy}, see Fig.~\ref{fig:spheres_illustration_3D_50}, with $64^3=$\numprint{262144} spherical inclusions, discretized by \numprint{1024}${}^3 \approx 1.07\times 10^9$ voxels, and computed the reference
\begin{figure}
	\begin{center}
		\includegraphics[width=.6\textwidth, trim = 530 90 440 250, clip]{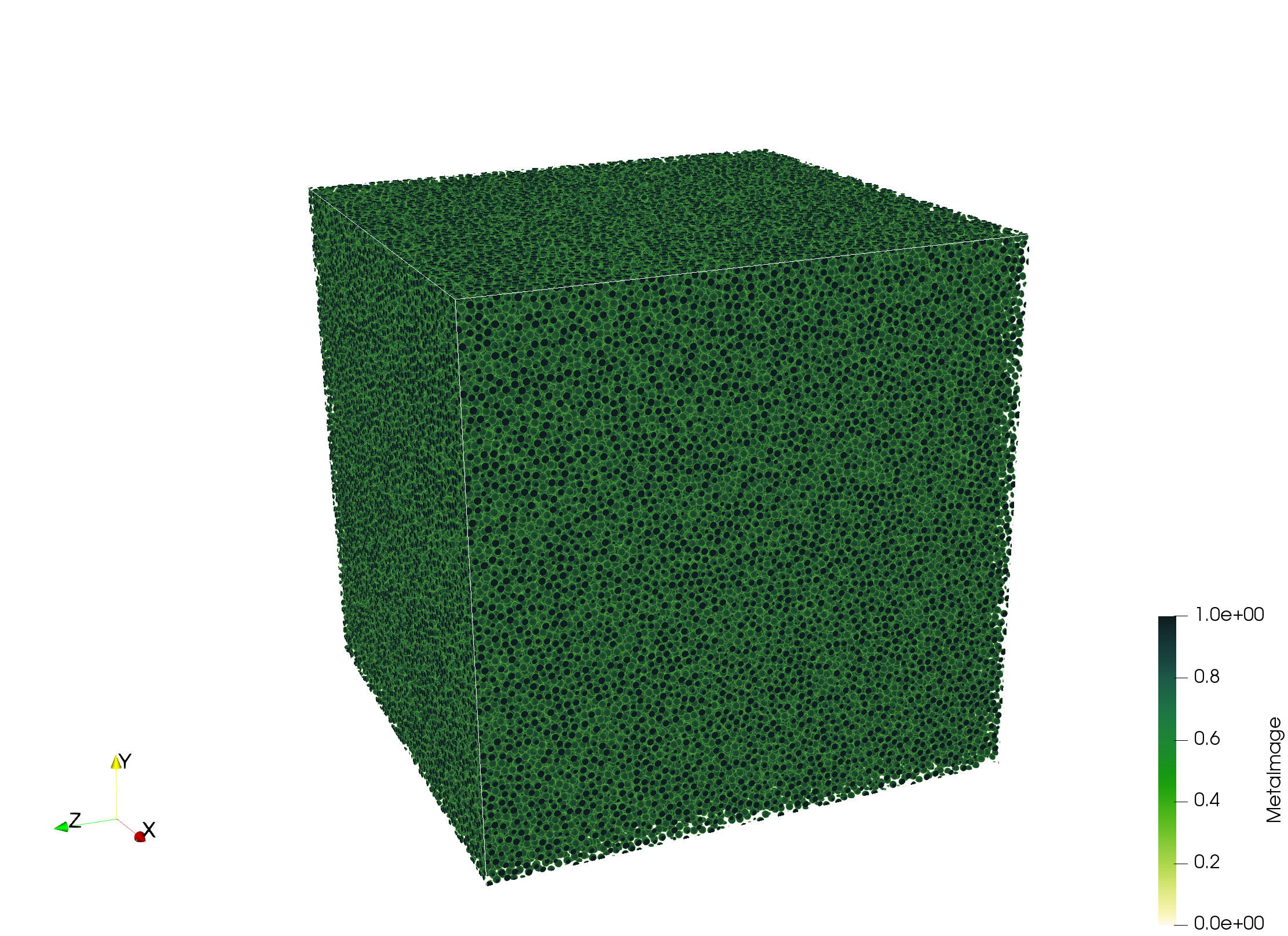}
	\end{center}
	\caption{A large volume element with $K=64$, i.e., $64^3 = \numprint{262144}$ spheres used for computing the reference effective conductivity \eqref{eq:ahom_ref_spheres}}
	\label{fig:spheres_illustration_3D_50}
\end{figure}
\begin{equation}\label{eq:ahom_ref_spheres}
	\bar{a} = 0.345228 \pm 0.000015 \, \text{W/(m$\cdot$K)}.
\end{equation}
Here, Student's $t$-distribution for $t=10$ is used for estimating a $99\%$ two-sided confidence interval based on the $(-1)$-shifted standard deviation computed for these ten samples.
\begin{table}[h!]
 \begin{center}
 \begin{tabular}{ll}
 \begin{tabular}{r|ccc}
 $K$	&periodized		& {snapshots}\\
  \hline
 $2$	& $0.345553\pm 5[-5]$	& $0.356449\pm 4[-4]$\\
 $4$	& $0.345538\pm 2[-5]$	& $0.351227\pm 2[-4]$\\
 $8$	& $0.345382\pm 6[-6]$	& $0.348215\pm 1[-4]$\\
 $16$	& $0.345291\pm 2[-6]$	& $0.346698\pm 5[-5]$\\
 \hline
 $64$	& $0.345224 \pm 1[-5]$	& --
 \end{tabular}
 &
 \begin{tabular}{r|ccc}
 $K$	& periodi{zed}		& {snapshots}\\
  \hline
 $2$	& $0.0017786$		& $0.0172593$\\
 $4$	& $0.0006614$		& $0.0080232$\\
 $8$	& $0.0002450$		& $0.0043529$\\
 $16$	& $0.0000871$		& $0.0020374$\\
 \hline
 $64$	& $0.0000107$		& --
 \end{tabular} 
 \end{tabular}
 \end{center}
 \caption{Computed effective conductivities $\bar{a}_L$ (left) and standard deviations (right) in W/(m$\cdot$K) for the spheres, computed using \numprint{10000} realizations}
 \label{tab:spheres_means_stdevs}
\end{table}
Tab.~\ref{tab:spheres_means_stdevs} contains the computed means and standard deviations for $K=2,4,8,16$ and the periodized/{snapshot} protocols. The mean values are supplemented by $99\%$-two sided confidence intervals, computed using the standard deviations, and rounded to the highest significant digit. Due to the large number of samples used, these confidence intervals are much tighter for $8 \le K \le 16$ than for the ten computations on large volume elements, cf. Fig.~\ref{fig:spheres_illustration_3D_50}.
\begin{figure}[h!]
 \begin{subfigure}{.49\textwidth}
  \includegraphics[width=\textwidth]{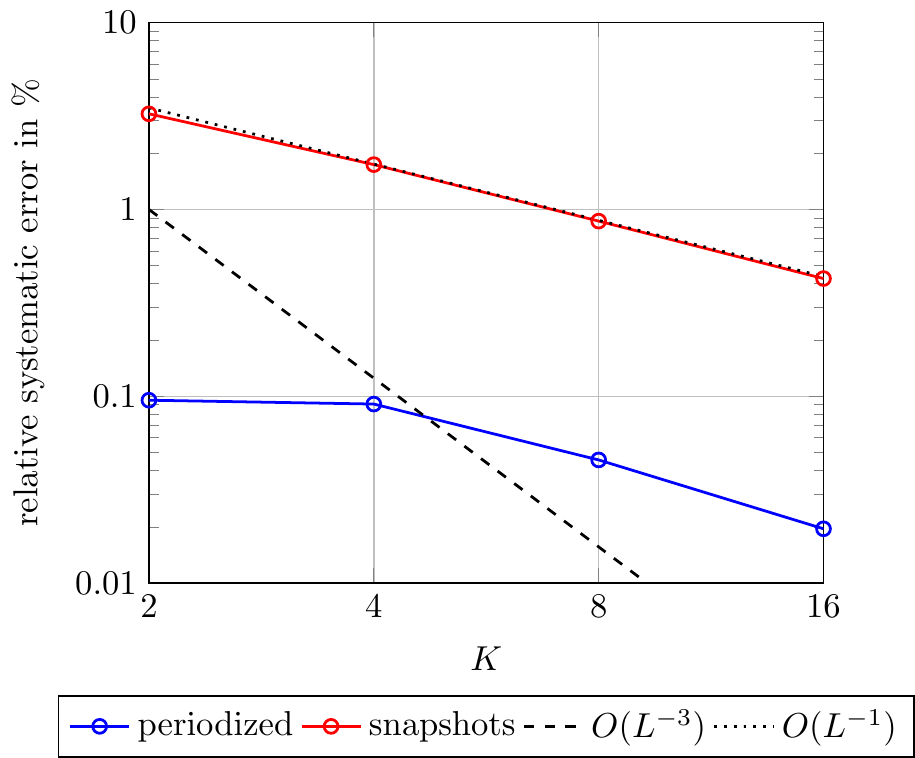}
  \caption{Relative systematic error $\left|\mean{a_L}\!/\bar{a} -1\right|$}
  \label{fig:spheres_mean}
 \end{subfigure}
  \begin{subfigure}{.49\textwidth}
  \includegraphics[width=\textwidth]{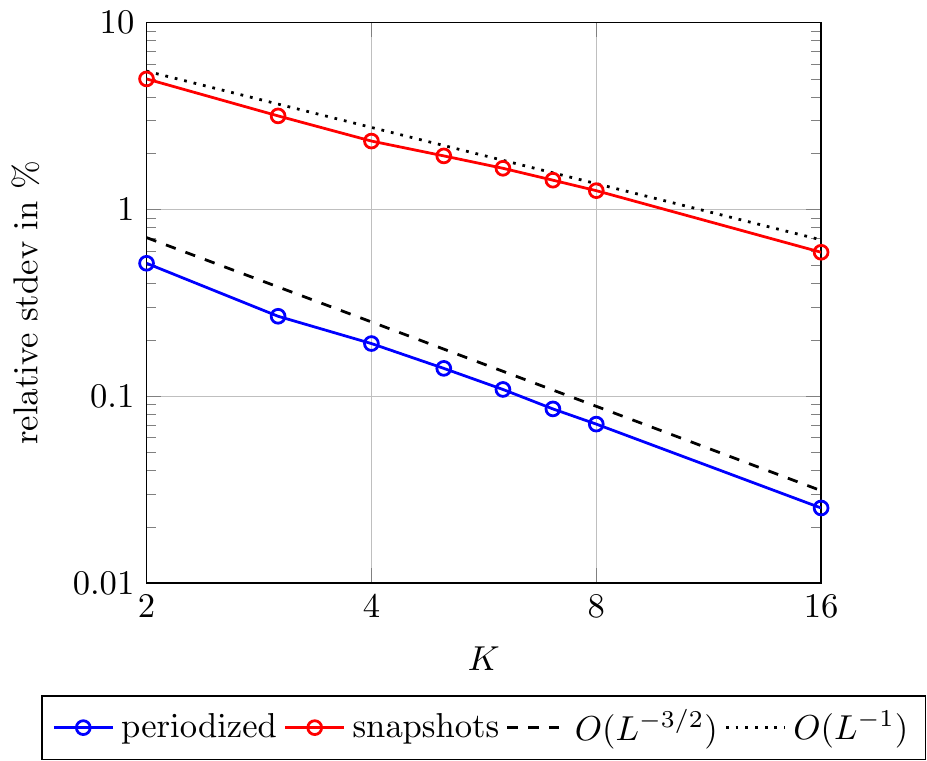}
  \caption{Relative standard deviation $\sqrt{\var{a_L}}/\mean{a_L}$}
  \label{fig:spheres_stdev}
 \end{subfigure}
 \caption{Convergence behavior of the systematic and the random error for spherical inclusions and the periodized/{snapshot} protocols, normalized according to equation \eqref{eq:ahom_ref_spheres}}
 \label{fig:spheres}
\end{figure}
The systematic error for the periodized ensembles is about an order of magnitude smaller than for their {snapshot} counterpart{s}, cf. Fig.~\ref{fig:spheres_mean}. For the latter, an $L^{-1}$-convergence behavior is evident.
However, from an engineering point of view, a systematic error below $0.1\%$ is sufficient - and this is computed using only $8$ (!) inclusions.\\
Taking a look at the relative standard deviation, cf. Fig.~\ref{fig:spheres_stdev}, we observe that the {snapshot-}sampling strategy leads to a convergence rate of $1/L$. In contrast, the standard deviation for the periodized ensemble decreases as $L^{-3/2}$. Also, there is a difference by an order of magnitude difference between the two protocols.\\
Due to proper scaling of the systematic and random errors in Fig.~\ref{fig:spheres}, we may also quantify the uncertainty involved in the computation. Indeed, the random error is about an order of magnitude larger than the systematic error for both protocols.\\
Mean and standard deviations only provide a limited amount of information concerning the full distribution of the random variable $a_L$. For this purpose, and from an engineering perspective, we may ask the following question: Suppose we run only a single computation, what is the probability of being $1\%$ ($0.1\%$) close to the reference \eqref{eq:ahom_ref_spheres}?
\begin{table}
 \begin{center}
 \begin{tabular}{ll}
 \begin{tabular}{l|rrr}
 $K$	& periodized	& {snapshots}\\
  \hline
 $2$	& $95.77$	& $12.20$\\
 $4$	& $100.00$	& $24.06$\\
 $8$	& $100.00$	& $45.45$\\
 $16$	& $100.00$	& $82.73$
 \end{tabular}
 &
 \begin{tabular}{l|rrr}
 $K$	& periodized	& {snapshots}\\
  \hline
 $2$	& $17.36$	& $1.42$\\
 $4$	& $35.29$	& $2.35$\\
 $8$	& $76.06$	& $4.68$\\
 $16$	& $99.93$	& $10.13$
 \end{tabular}
 \end{tabular} 
 \end{center}
 \caption{Empirical probability (in $\%$, \numprint{10000} realizations) of being $1\%$-close (left) and $0.1\%$-close (right) to $\bar{a}$ \eqref{eq:ahom_ref_spheres} for spherical fillers and $30\%$ volume fraction}
 \label{tab:spheres_success_probability}
\end{table}
Using the \numprint{10000} computations for each setup, the empirical probabilities are collected in Tab.~\ref{tab:spheres_success_probability}. We see that, for the periodized protocol, more than two correct digits are computed with a chance of more than $95\%$ using $2^3 = 8$ spheres. For higher $K$, each of the \numprint{10000} computations predicted two significant digits correctly. In contrast, for the {snapshot} protocol, even using $16^3 = 4096$ spheres led to a lower success probability than for eight spheres {in the periodized setting}.\\
Computing the third significant digit with a single run is more challenging, and $K\ge 8$ is required for the periodized protocol to have a higher success than failure probability. Only for $K=16$, i.e., \numprint{4096} spheres, three significant digits may be computed almost surely. Indeed, only $7$ out of \numprint{10000} runs failed to provide three significant digits.\\
In contrast, for the {snapshot} approach, the success probability is about an order of magnitude smaller.


%
%
%
%

\subsection{Circular inclusions}
\label{sec:results_disks}

Subsequent to studying spherical inclusions, we turn to circular inclusions in two spatial dimensions. Microstructures of this type serve as models for continuous fibers, i.e., parallel cylindrical inclusions whose length is much larger than the diameter, and the effective behavior may be determined from a two-dimensional model. In turn, we may investigate the changes in convergence rates for stochastic homogenization when decreasing the ambient dimension.\\
We followed a similar protocol as for the spherical case, with the notable difference that we only need to cut out $1/4$th from a sample containing $(2K)^2$ disks to obtain the {snapshot}. To ensure some degree of compatibility, we retained the {area}-fraction steps and the isolation distance for the MCM algorithm. We also used $(16K)^2$ pixels for discretizing the geometries filled by $K^2$ disks, on average. Due to the two-dimensional situation, we were able to treat larger $K$ than for the three-dimensional case. 
\begin{figure}
	\begin{subfigure}{.49\textwidth}
		\includegraphics[width=\textwidth, trim = 600 300 600 300, clip]{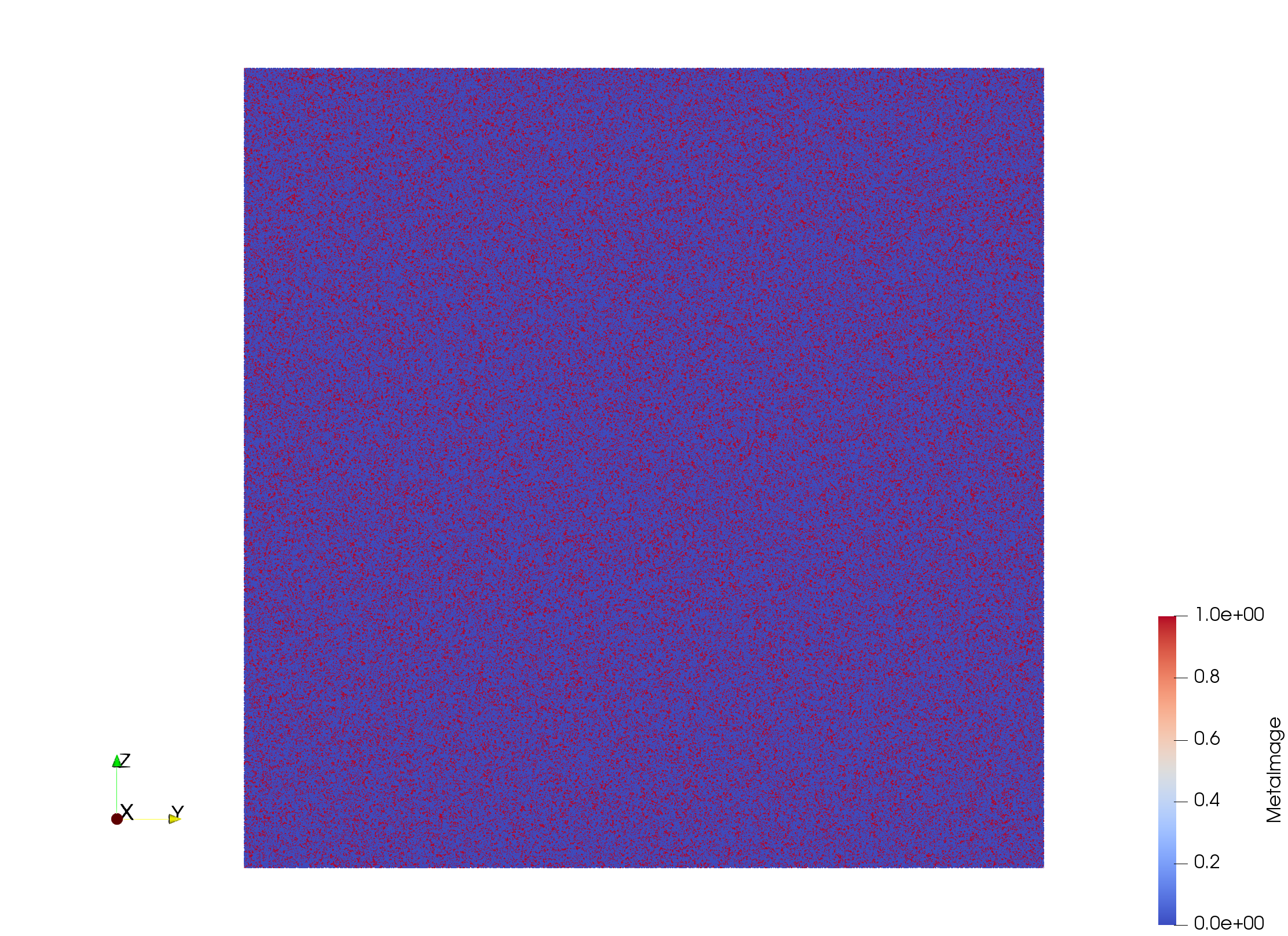}
		\caption{Total view (\numprint{16384}${}^2$ pixels)}
		\label{fig:disks_ref_total}
	\end{subfigure}
	\begin{subfigure}{.49\textwidth}
		\includegraphics[width=\textwidth, trim = 600 300 600 300, clip]{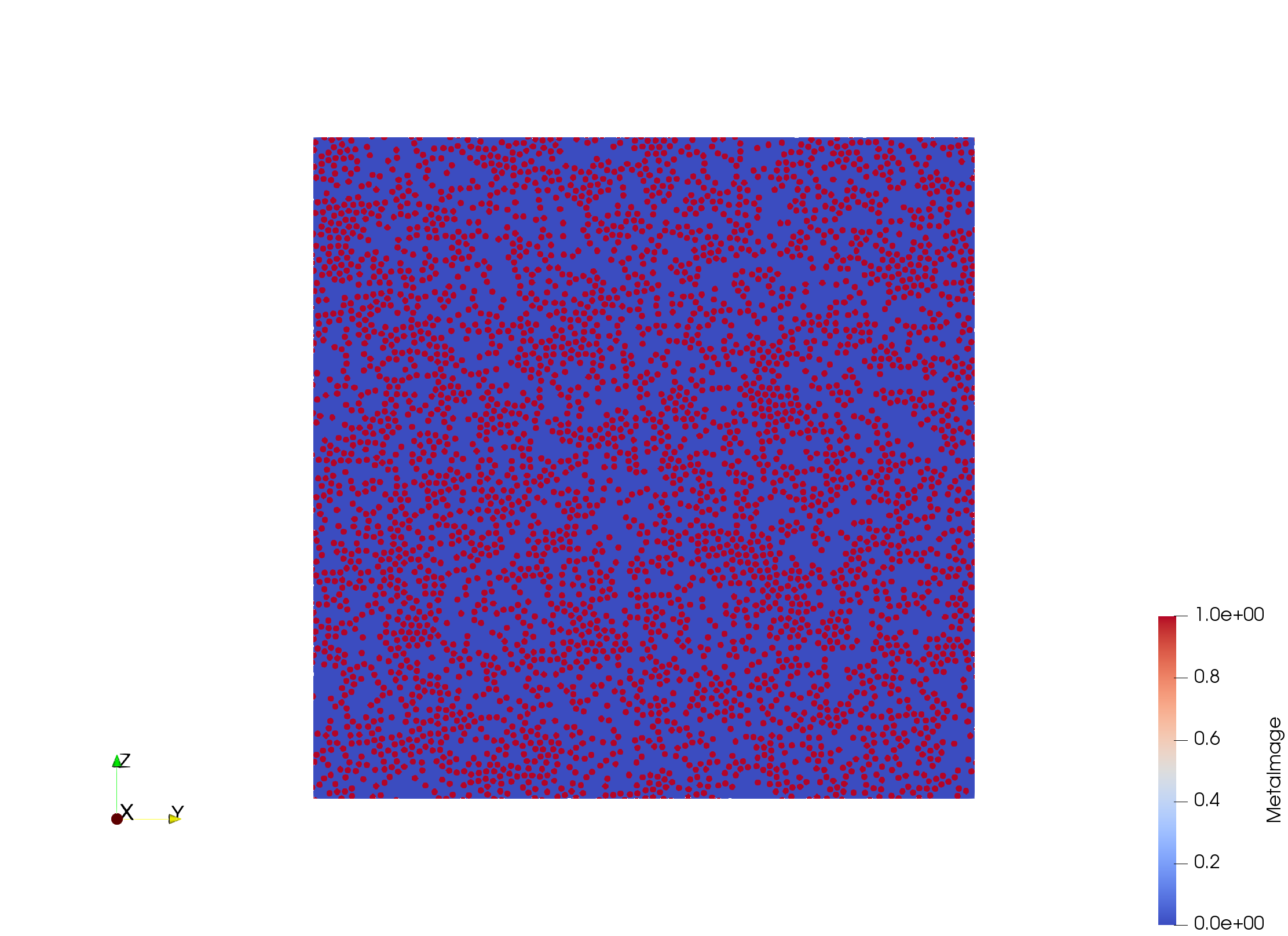}
		\caption{Cut-out of $(1/16)\times (1/16)$th (\numprint{1024}${}^2$ pixels)}
		\label{fig:disks_ref_cut}
	\end{subfigure}	
	\caption{First of the ten microstructures with \numprint{1024}${}^2$ disks used for computing the effective conductivity \eqref{eq:ahom_ref_disks}}
	\label{fig:disks_ref}
\end{figure}
For reference, we generated ten microstructures {with a periodized protocol} with \numprint{1024}${}^2=$\numprint{1048576} circular inclusions, see Fig.~\ref{fig:disks_ref}, discretized by \numprint{16384}${}^2$ pixels. We obtained a reference conductivity, measured transverse to the fibers, of
\begin{equation}\label{eq:ahom_ref_disks}
	\bar{a} = 0.3174257 \pm 0.0000214 \, \text{W/(m$\cdot$K)},
\end{equation}
including a two-sided $99\%$-confidence interval based on Student's $t$-distribution for the ten drawn samples, and using identical material parameters as for the spherical inclusions. Notice that the effective conductivity \eqref{eq:ahom_ref_disks} transverse to continuous fibers is smaller than the effective conductivity of spherical fillers \eqref{eq:ahom_ref_spheres} at the same {filler} fraction.\\
To assess the convergence behavior of the transverse effective conductivity,
\begin{table}[h!]
 \begin{center}
 \begin{tabular}{r|cc}
 $K$	&periodized		& {snapshots}\\
  \hline
  $2$ & $0.316286\pm 2[-4]$ & $0.324902\pm 1[-3]$\\
  $4$ & $0.317640\pm 1[-4]$ & $0.320024\pm 7[-4]$\\
  $8$ & $0.317517\pm 6[-5]$ & $0.318881\pm 4[-4]$\\
 $16$ & $0.317505\pm 3[-5]$ & $0.318172\pm 2[-4]$\\
 $32$ & $0.317459\pm 2[-5]$ & $0.317617\pm 1[-4]$\\
 $64$ & $0.317438\pm 7[-6]$ & $0.317591\pm 5[-5]$\\
 \hline
 \numprint{1024}& $0.317426\pm 2[-5]$& --\\
 \end{tabular}
 \begin{tabular}{r|cc}
 $K$	&periodized		& {snapshots}\\
  \hline
  $2$ & $0.007254$ & $0.045318$\\
  $4$ & $0.004890$ & $0.026951$\\
  $8$ & $0.002419$ & $0.014757$\\
 $16$ & $0.001181$ & $0.007808$\\
 $32$ & $0.000592$ & $0.004039$\\
 $64$ & $0.000291$ & $0.002059$\\
 \hline
 \numprint{1024}& $0.000021$ & --\\
 \end{tabular} 
 \end{center}
 \caption{Computed effective conductivities $\bar{a}_L$ (left) and standard deviations (right) in W/(m$\cdot$K) for the circular inclusions, computed using \numprint{10000} realizations}
 \label{tab:disks_means_stdevs}
\end{table}
we generated \numprint{10000} microstructures for $K$ ranging from $2$ to $64$ in dyadic steps, both for the periodized and the {snapshot} protocol, and computed the $\bar{a}$-value. The results are collected in Tab.~\ref{tab:disks_means_stdevs}, together with the estimated standard deviations. For comparison, also the computed value for $K=$\numprint{1024} was included. As for the spherical case, the confidence interval for the $K=1024$ reference computation is comparatively large, approximately identical to the $K=32$ case. This effect is a consequence of using $10$ samples for $K=$\numprint{1024} instead of \numprint{10000} samples for $K=32$. Also notice that $K=64$ corresponds to $K^2 = $\numprint{4096} circular inclusions, discretized on a regular grid with \numprint{1024}${}^2$ pixels.
\begin{figure}[h!]
 \begin{subfigure}{.49\textwidth}
  \includegraphics[width=\textwidth]{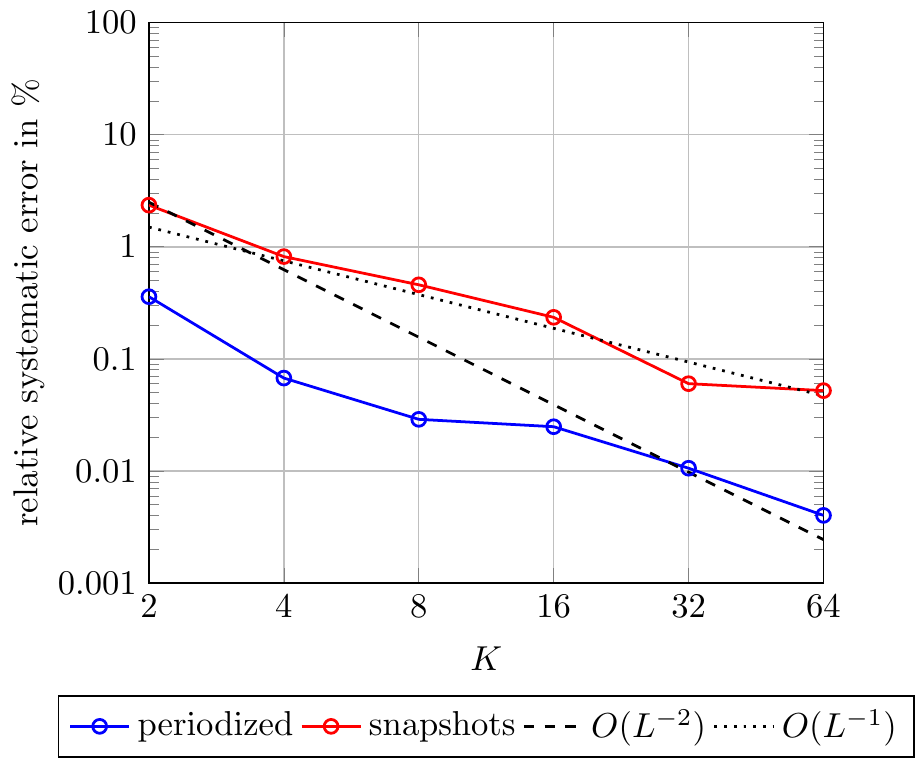}
  \caption{Relative systematic error $\left|\mean{a_L}\!/\bar{a} -1\right|$}
  \label{fig:disks_mean}
 \end{subfigure}
  \begin{subfigure}{.49\textwidth}
  \includegraphics[width=\textwidth]{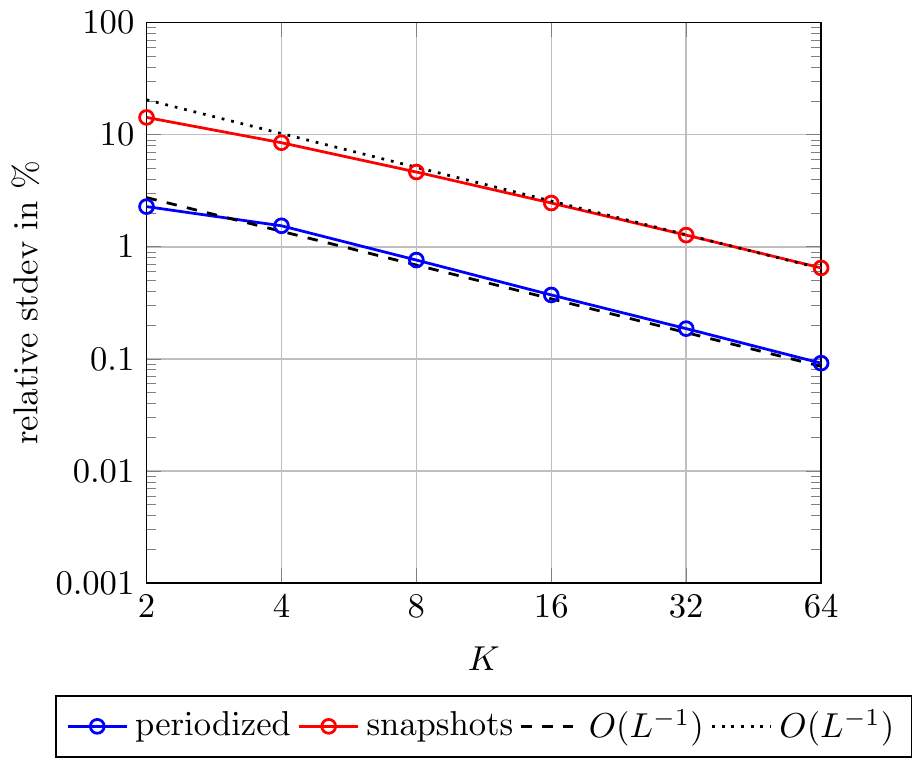}
  \caption{Relative standard deviation $\sqrt{\var{a_L}}/\mean{a_L}$}
  \label{fig:disks_stdev}
 \end{subfigure}
 \caption{Convergence behavior of the systematic and the random error for circular inclusions and the periodized/{snapshot} protocols, normalized according to equation \eqref{eq:ahom_ref_disks}}
 \label{fig:disks}
\end{figure}
We observe, see Fig.~\ref{fig:disks_mean}, that the systematic error of the periodized protocol is consistently smaller than for the {snapshot} approach, almost by an order of magnitude. {For the {snapshot} ensemble,} the systematic error follows an $1/L$-rate, as predicted. Evaluating the periodized protocol is much more difficult, as the systematic error is extremely small. In particular, and in view of the confidence interval of the reference computation, see Tab.~\ref{tab:disks_means_stdevs}, we cannot safely provide a convergence rate.
For the standard deviation, see Fig.~\ref{fig:disks_stdev}, both protocols lead to an $1/L$-convergence rate. Thus, in contrast to the spherical case, there is no difference in convergence rates between the periodized and the {snapshot} ensemble. We will discuss the reasons behind this effect in {Section} \ref{sec:discussion}.
\begin{table}
 \begin{center}
 \begin{tabular}{cc}
 \begin{tabular}{l|rr}
 $K$	& periodized	& {snapshots}\\
  \hline
 $2$    & $ 32.35$ 	& $ 5.61$\\
 $4$    & $ 50.27$ 	& $ 9.47$\\
 $8$    & $ 80.93$ 	& $17.52$\\
 $16$	& $ 99.25$	& $30.95$\\
 $32$	& $100.00$	& $56.77$\\
 $64$	& $100.00$	& $87.47$\\
 \end{tabular}
 &
  \begin{tabular}{l|rr}
 $K$	& periodized	& {snapshots}\\
  \hline
 $2$    & $ 3.47$ 	& $ 0.67$\\
 $4$    & $ 5.47$ 	& $ 0.92$\\
 $8$    & $10.81$ 	& $ 1.78$\\
 $16$	& $21.30$	& $ 3.24$\\
 $32$	& $40.60$	& $ 5.86$\\
 $64$	& $72.71$	& $12.31$\\
 \end{tabular}
 \end{tabular}
 \end{center}
 \caption{Empirical probability (in $\%$, \numprint{10000} realizations) of being $1\%$-close (left) and $0.1\%$-close (right) to $\bar{a}$ \eqref{eq:ahom_ref_disks} for circular inclusions}
 \label{tab:disks_success_probability}
\end{table}
The standard deviation for the {snapshot} approach is about an order of magnitude larger than for the periodized setting. Comparing the levels of stochastic and systematic error, see \ref{fig:disks}, for both approaches the random error is about an order of magnitude larger than the systematic error.\\
As for the spherical inclusions treated in {Section} \ref{sec:results_spheres}, we take a look at the empirical success probabilities of obtaining two and three, respectively, significant digits with only a single computation, see Tab.~\ref{tab:disks_success_probability}. For the periodized protocol, $16^2$ circular inclusions are required to obtain two correct digits with $99\%$ probability, i.e., more than $100$ fibers are required. In contrast, for the {snapshot} {strategy}, $32^2 =$\numprint{1024} inclusions are required to exceed $50\%$ success probability, a value that is exceeded for $4^2 = 16$ inclusions in case of the periodized protocol.\\
Obtaining three correct significant digits is difficult in the two-dimensional setting, even for the periodized protocol. Indeed, even working with $64^2=$\numprint{4096} inclusions leads to a failure rate of about $30\%$.

%
%

%
%
%


\subsection{Short-fiber reinforced composites}
\label{sec:results_fibers}

\begin{figure}
	\begin{subfigure}[t]{.171\textwidth}
		\includegraphics[width=\textwidth, trim = 490 210 1319 997, clip]{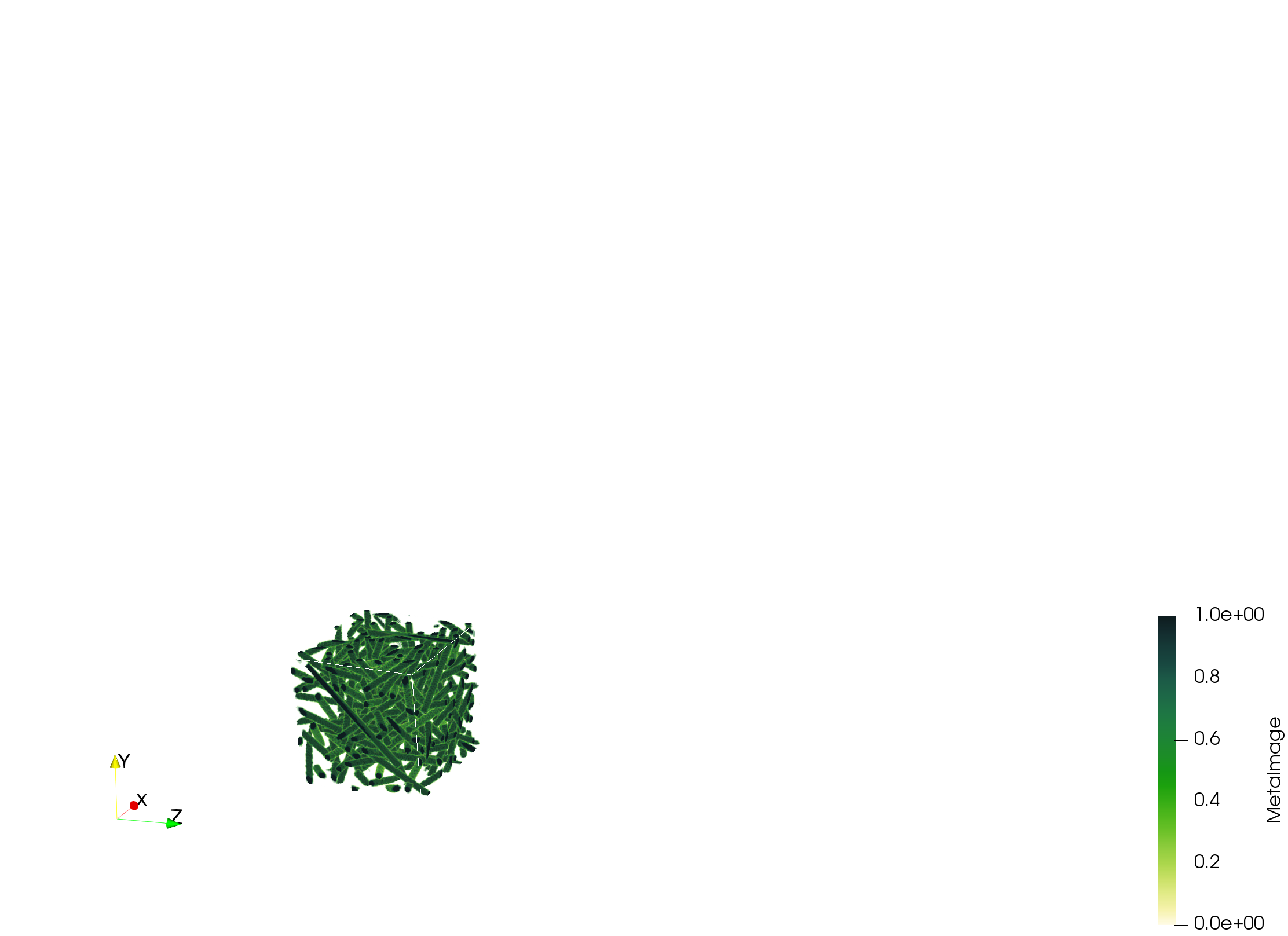}
		\caption{$L=\ell$, $77$ fibers}
		\label{fig:fibers_illustration_3D_1}
	\end{subfigure}
	\begin{subfigure}[t]{.307\textwidth}
		\includegraphics[width=\textwidth, trim = 470 190 1008 690, clip]{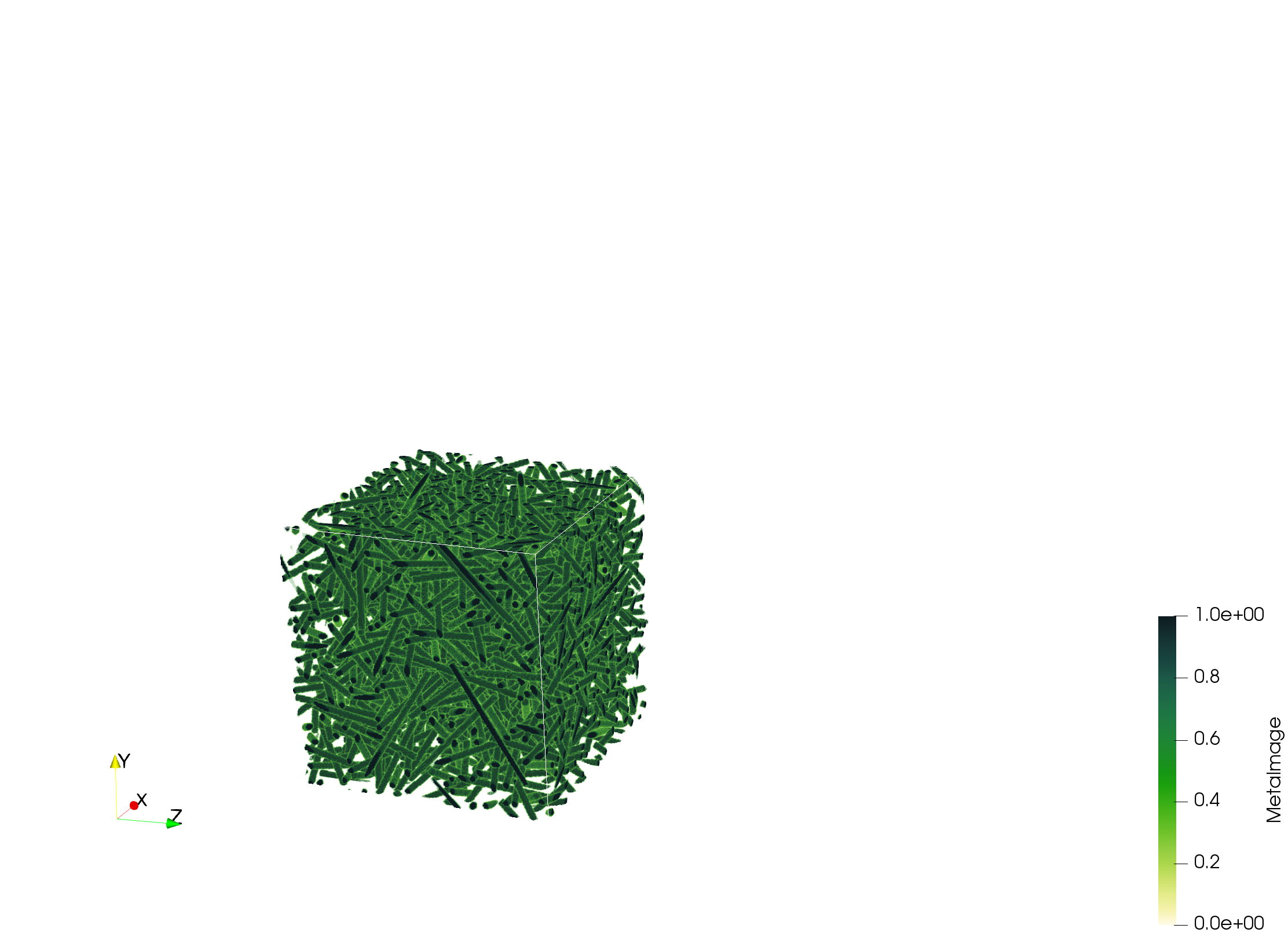}
		\caption{$L=2\ell$, $612$ fibers}
		\label{fig:fibers_illustration_3D_2}
	\end{subfigure}
	\begin{subfigure}[t]{.512\textwidth}
		\includegraphics[width=\textwidth, trim = 430 110 550 280, clip]{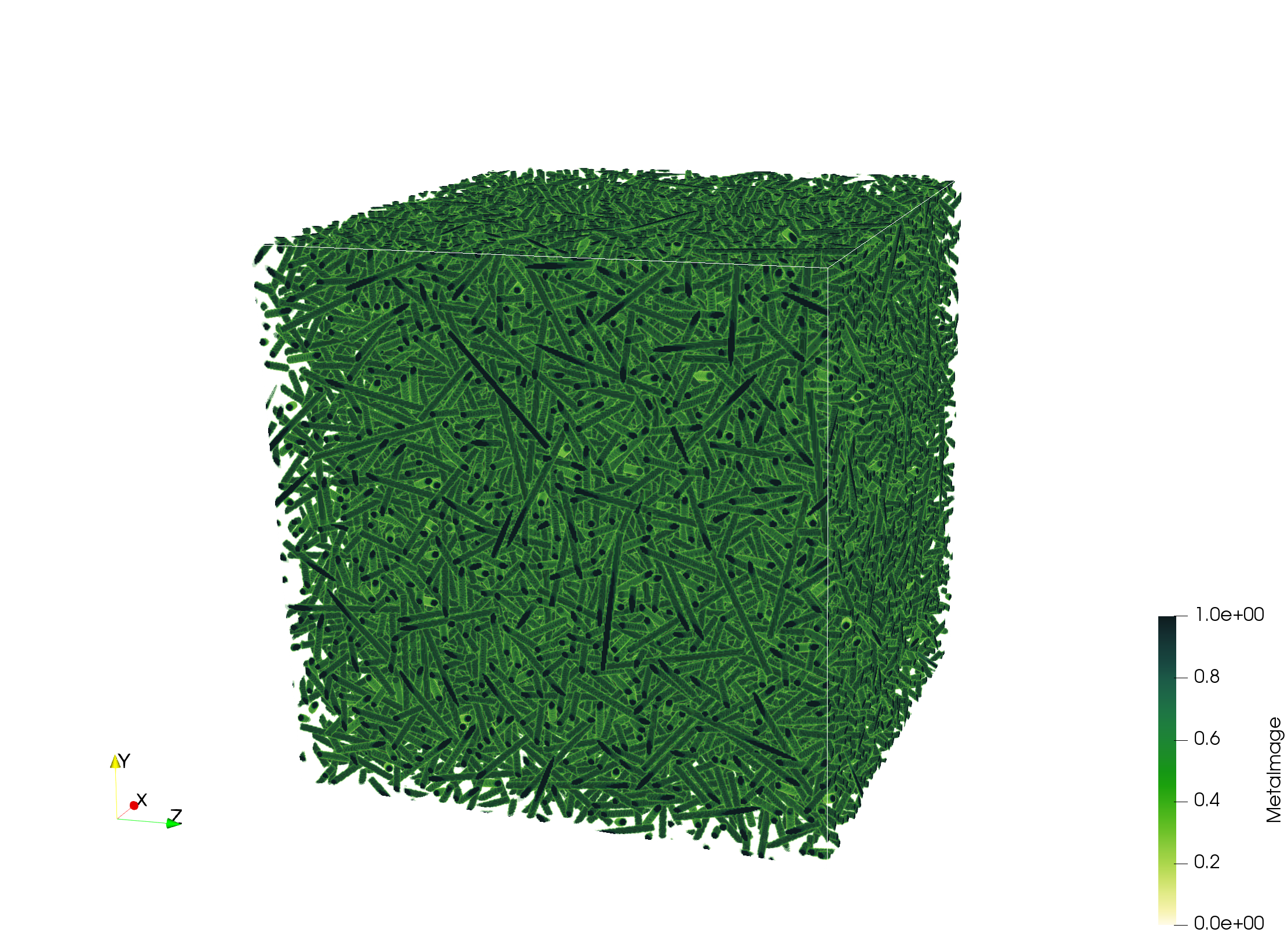}
		\caption{$L=4\ell$, \numprint{4890} fibers}
		\label{fig:fibers_illustration_3D_4}
	\end{subfigure}
	\caption{Periodically generated short-fiber reinforced elements with increasing fiber count}
	\label{fig:fibers_illustration_3D}
\end{figure}
Last but not least, we consider short-fiber reinforced composites, i.e., microstructures consisting of non-overlapping cylindrical inclusions. More precisely, we consider identically shaped  cylindrical fibers with length $\ell$ and diameter $d$ at an aspect ratio $\ell/d$ of $20$, with an isotropic fiber orientation, encoded by an isotropic fourth-order fiber-orientation tensor~\cite{AdvaniTucker}.\\
We shall also consider a periodized and a {snapshot} protocol. In the periodized setting, we fill to $15\%$ volume in three $5\%$-steps by Sequential Addition and Migration~\cite{SAM}, see {Section} \ref{sec:microstructure_gen_fibers}, with a minimum distance of $20\%$ of the fiber's diameter, and consider cubic volume elements $Q_L$, s.t. the edge length $L$ is an integer multiple of the fiber length $\ell$, i.e., $L/\ell = 1,2,3,4$, see Fig.~\ref{fig:fibers_illustration_3D} for examples.\\
For the {snapshot} protocol, we generate cells $Q_{\tfrac{3}{2}\, L}$ {with the periodic protocol} und extract a subcell of dimension $L^3$. 
As in the previous two {Section}s, we endow the matrix with the (isotropic) conductivity parameters of PP ($0.2 \text{W/(m$\cdot$K)}$) and the fibers with those of E-glass ($1.2 \text{W/(m$\cdot$K)}$). 
\begin{figure}
	\begin{center}
		\includegraphics[width=.6\textwidth, trim = 430 110 550 280, clip]{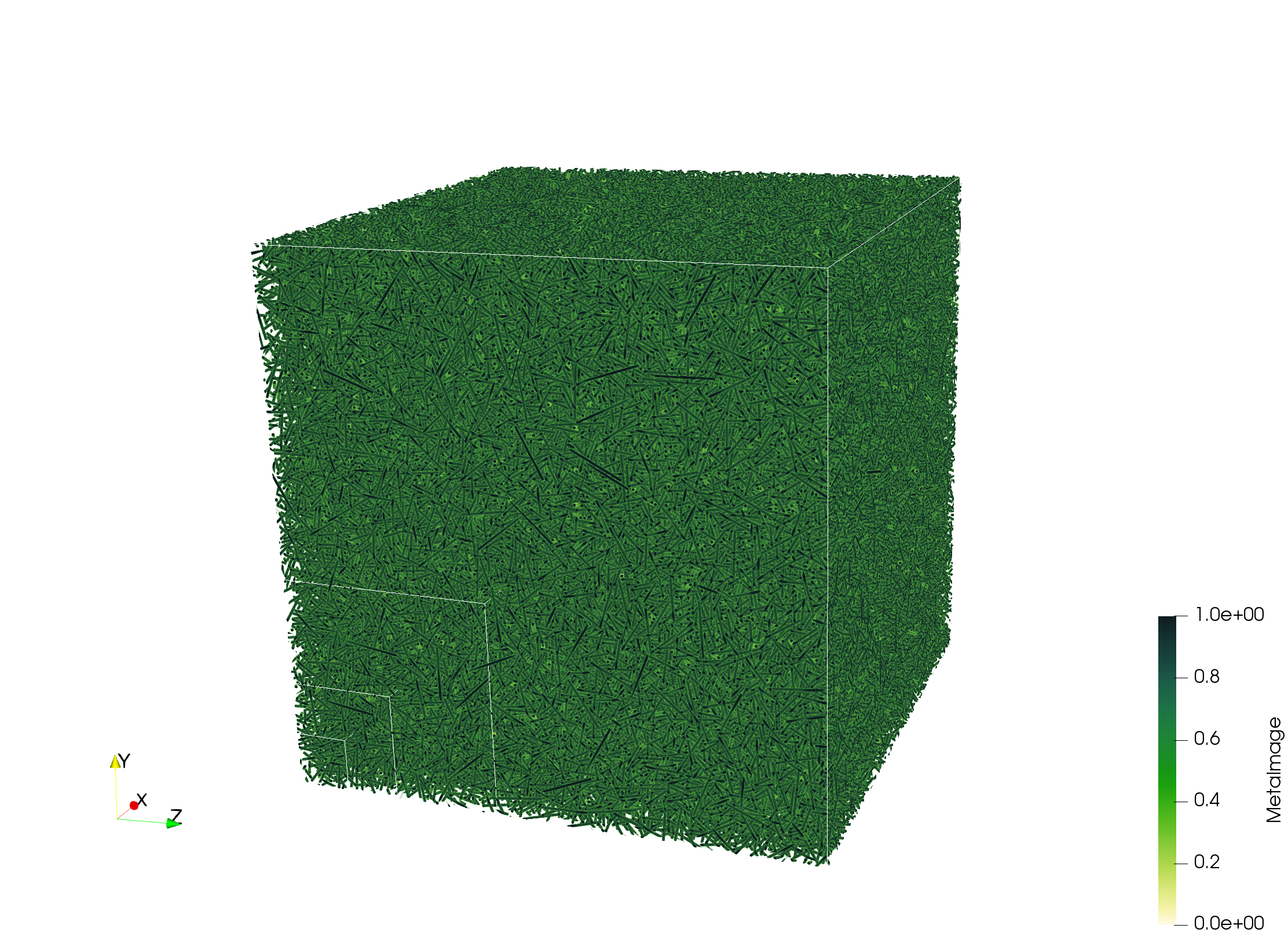}
	\end{center}
	\caption{Large volume element with $L/\ell=10$, including \numprint{76397} fibers. Smaller cell sizes with $L/\ell \in \{1,2,4\}$ are indicated by white boxes in the lower left corner}
	\label{fig:fibers_illustration_3D_large}
\end{figure}
Each fiber is discretized by $5$ voxels per diameter~\cite{MuellerKabelEtAl2015}, i.e., $100$ voxels per length $\ell$. This translates into $(L/\ell \times 100)^3$ voxels in total, i.e., $100^3$ voxels for $L=\ell$ and $400^3$ voxels for $L = 4\ell$.\\
We use the average of $10$ computed effective conductivities for samples with $L/\ell = 10$ (\numprint{1000}${}^3$ voxels, see Fig.~\ref{fig:fibers_illustration_3D_large}) as the reference: 
\begin{equation}\label{eq:ahom_ref_fibers}
	\bar{a} = 0.281008 \pm 0.0000055 \,\text{W/(m$\cdot$K)},
\end{equation}
including a $99\%$ two-sided confidence interval, as before.
\begin{table}[h!]
 \begin{center}
 \begin{tabular}{ll}
 	\begin{tabular}{r|cc}
 	$\frac{L}{\ell}$	&periodized		& {snapshots}\\
  	\hline
	$1$ & $0.281079 \pm 6[-6]$ & $0.280321 \pm 1[-4]$\\
	$2$ & $0.281211 \pm 2[-6]$ & $0.280583 \pm 5[-5]$\\
	$3$ & $0.281060 \pm 1[-6]$ & $0.280716 \pm 3[-5]$\\
	$4$ & $0.281042 \pm 1[-6]$ & $0.280766 \pm 2[-5]$\\
	\hline
	$10$ & $0.281008 \pm 6[-6]$& --\\
 	\end{tabular}
 	&
 	\begin{tabular}{r|cc}
 	$\frac{L}{\ell}$	& periodized		& {snapshots}\\
  	\hline
 	$1$	& $0.000223$ & $0.003678$\\
	$2$	& $0.000087$ & $0.001823$\\
	$3$	& $0.000047$ & $0.001178$\\
	$4$	& $0.000031$ & $0.000837$\\ 	
	\hline 
	$10$& $0.000006$ & --
	\end{tabular} 
 \end{tabular}
 \end{center}
 \caption{Computed effective conductivities $\bar{a}_L$ (left) and standard deviations (right) in W/(m$\cdot$K) for short fibers, computed using \numprint{10000} realizations}
 \label{tab:fibers_means_stdevs}
\end{table}
The empirical means of \numprint{10000} computed effective conductivities for the considered volume-element sizes and the two protocols are listed in Tab.~\ref{tab:fibers_means_stdevs}, together with two-sided $99\%$ confidence intervals and standard deviations. For completeness, the reference data \eqref{eq:ahom_ref_fibers} are also included.\\
Notice that the confidence intervals are very tight in this setting. Indeed, due to the shape of the inclusions, even for $L = \ell$, twenty fiber diameters fit across an edge of such a volume element. In fact, the confidence interval for $L = \ell$ is very close to the confidence interval for $L = 10 \, \ell$, as \numprint{10000} samples were evaluated for the former, and only ten samples for the latter. Even for the {snapshot-}sampling strategy the {random} errors are small. 
\begin{figure}
 \begin{subfigure}{.49\textwidth}
  \includegraphics[width=\textwidth]{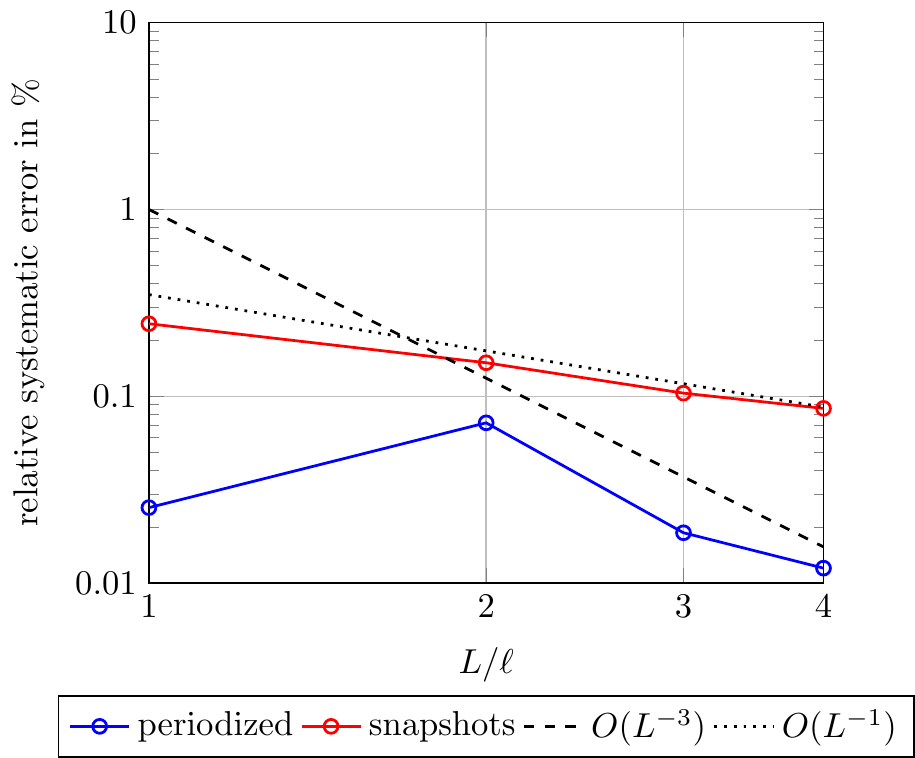}
  \caption{Relative systematic error $\left|\mean{a_L}\!/\bar{a} -1\right|$}
  \label{fig:fibers_mean}
 \end{subfigure}
  \begin{subfigure}{.49\textwidth}
  \includegraphics[width=\textwidth]{figs/fig11a}
  \caption{Relative standard deviation $\sqrt{\var{a_L}}/\mean{a_L}$}
  \label{fig:fibers_stdev}
 \end{subfigure}
 \caption{Convergence behavior of the systematic and the random error for short fibers and the periodized/{snapshot} protocols, normalized according to equation \eqref{eq:ahom_ref_fibers}}
 \label{fig:fibers}
\end{figure}
This accuracy is also reflected in the normalized systematic and random errors, see Fig.~\ref{fig:fibers_stdev}. For all settings considered, they do not exceed $2\%$.\\
For the systematic error, see Fig.~\ref{fig:fibers_mean}, the {snapshot} strategy shows the already familiar $1/L$-scaling, whereas the periodized sampling protocol appears to scale as $L^{-3}$ for $L\ge 2 \, \ell$. Between $L=\ell$ and $L = 3 \, \ell$, the systematic error does not decrease monotonically, presumably due to a geometric influence. Indeed, for $L = \ell$, an edge of the volume element under consideration is exactly equal to a fiber length. The comparatively low systematic error for this case may thus be a results of the severe restrictions in fiber arrangement for $L = \ell$. In magnitude, the systematic error for the {snapshot} approach is about half an order of magnitude larger than for the periodized sampling.\\
For the random error, see Fig.~\ref{fig:fibers_stdev}, we observe an $L^{-1}$ and $L^{-\frac{3}{2}}$ scaling for the {snapshot} and periodized protocol, respectively, in agreement with the observations made for spheres, see Fig.~\ref{fig:spheres}. Also, there is about an order of magnitude difference in the absolute standard deviations for both approaches, already evident in Tab.~\ref{tab:fibers_means_stdevs}. It is interesting to note that, for the periodized sampling, the random error is on the same order of magnitude as the systematic error. This behavior contrasts what we observed for spherical and circular inclusions.\\
As for the other inclusion types, the empirical probabilities for remaining within $1\%$ (and $0.1\%$) relative error for only a single computation are given in Tab.~\ref{tab:fibers_success_probability}.
\begin{table}
 \begin{center}
 \begin{tabular}{ll}
 \begin{tabular}{l|rr}
 $\frac{L}{\ell}$	& periodized	& {snapshots}\\
  \hline
 $1$	& $100.00$	& $55.09$\\
 $2$	& $100.00$	& $86.54$\\
 $3$	& $100.00$	& $98.02$\\
 $4$	& $100.00$	& $99.91$\\
 \end{tabular}
 &
 \begin{tabular}{l|rr}
 $\frac{L}{\ell}$	& periodized	& {snapshots}\\
  \hline
 $1$	 & $77.12$	&  $6.14$\\
 $2$	 & $81.91$	& $11.63$\\
 $3$	& $100.00$	& $17.85$\\
 $4$	& $100.00$	& $24.59$\\
 \end{tabular}
 \end{tabular} 
 \end{center}
 \caption{Empirical probability (in $\%$, \numprint{10000} realizations) of being $1\%$-close (left) and $0.1\%$-close (right) to $\bar{a}$ \eqref{eq:ahom_ref_spheres} for fibers}
 \label{tab:fibers_success_probability}
\end{table}
For the periodized sampling strategy, each of the \numprint{40000} computation was correct to two significant digits. Also, the chance for obtaining three correct digits exceeds $3/4$ for $L = \ell$, already. For $L = 3\,\ell$ and above, each individual computation was guaranteed to feature three correct digits. In contrast, for the {snapshot} sampling, the success probabilities are much smaller. An edge length $L = 3 \, \ell$ is required to obtain two correct digits with $98\%$ probability, and the involved effort is at least $3^3 = 27$ times as high as using {an element from the periodized ensemble} and $L = \ell$. Obtaining higher accuracy than two digits appears out of reach for the {snapshot} strategy. Even for $L = 4\,\ell$, only a quarter of the \numprint{10000} computations led to results with three correct leading digits.


%
%
%
%


\section{On the non-standard scaling of the random error in three dimensions}
\label{sec:discussion}

This section is intended to complement the results of the previous section. More precisely, we provide insights into the difference in the decay behavior of the random error for the {snapshot} ensembles compared to their periodized counterparts. It will also become clear why we observe this behavior in three spatial dimensions only.\\
For a random conducting medium with two distinct (positive) conductivities $\alpha_1$ and $\alpha_2$, a series expansion in terms of the material-contrast {related quantity} 
\[
	{\rho = \frac{ \sqrt{\alpha_1} - \sqrt{\alpha_2}  }{ \sqrt{\alpha_1} + \sqrt{\alpha_2} }  }
\]
 permits identifying the leading-order term {in the estimate}
\begin{equation}\label{eq:stochastic_error_two_phase_isotropic}
	{\sqrt{\mean{\left\|\bar{A} - \bar{A}_L\right\|^2}} \le 4 \sqrt{\alpha_1 \alpha_2} \sqrt{\var{\phi_L}} \, \rho + O(\rho^2)}.
\end{equation}
We refer to Appendix \ref{sec:theory_series} for a self-contained derivation. Thus, the decay rate of the random error is closely tied to the decay rate of the variance of the sampled volume fraction. In Fig.~\ref{fig:phi_convergence_spheres}, we compare the standard deviations of the volume fractions $\phi_L$ for the periodized and the {snapshot} protocol and spherical fillers.
\begin{figure}
	\begin{subfigure}{.53\textwidth}
		\includegraphics[width=\textwidth]{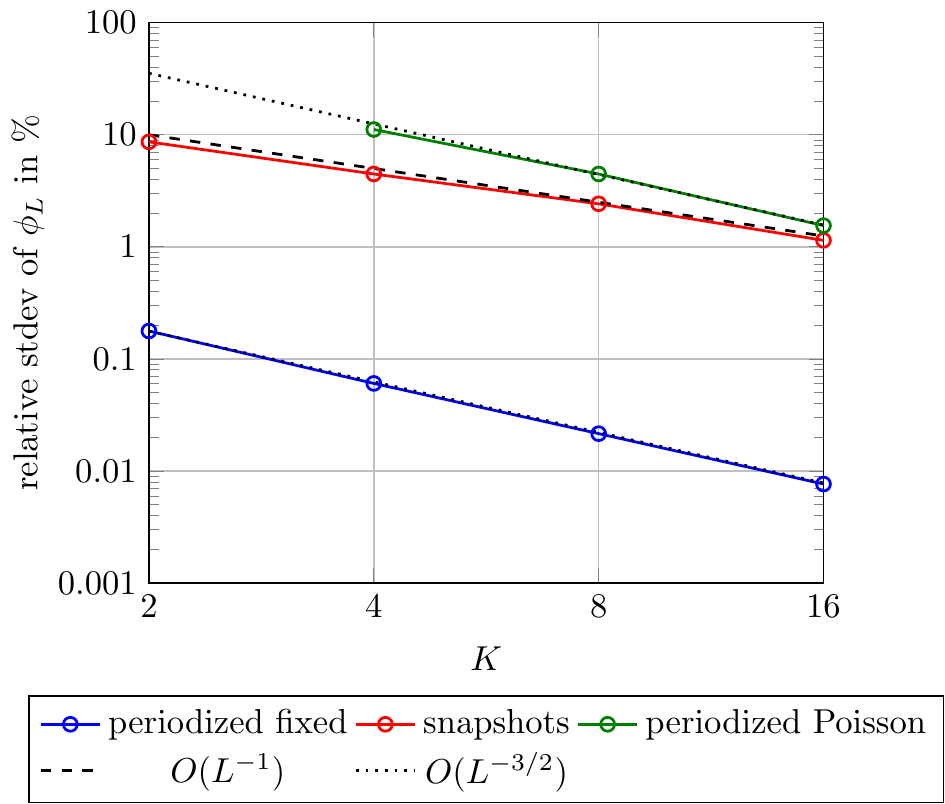}
		\caption{Spherical inclusions in $3$D}
		\label{fig:phi_convergence_spheres}
	\end{subfigure}
	\begin{subfigure}{.45\textwidth}
		\includegraphics[width=\textwidth]{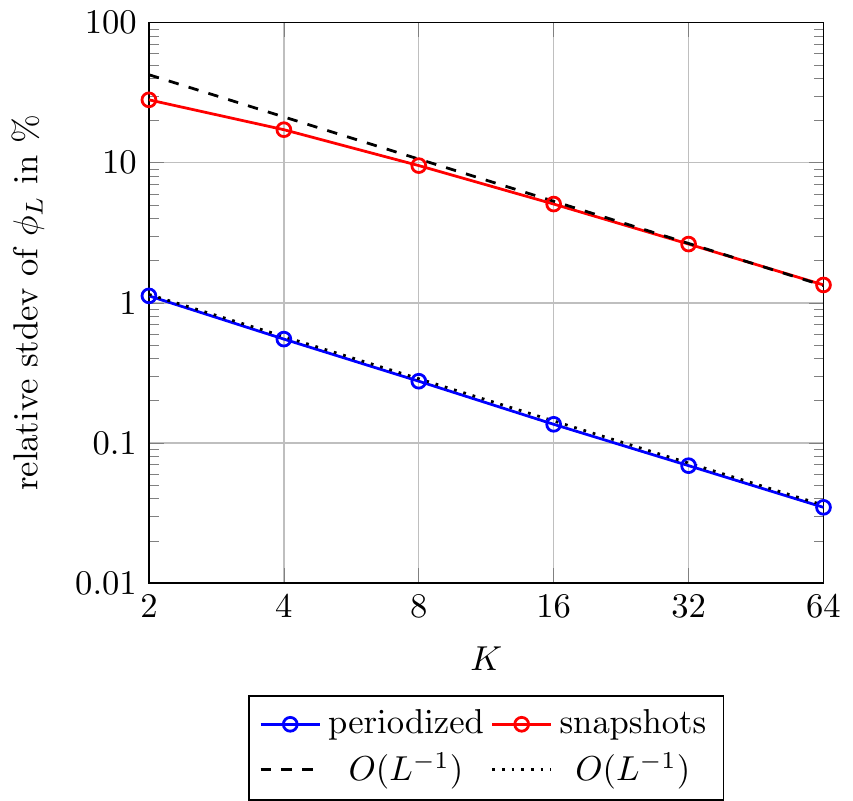}
		\caption{Circular inclusions in $2$D}
		\label{fig:phi_convergence_disks}
	\end{subfigure}
	\caption{Convergence behavior of the standard deviation for computing the volume fraction $\phi_L$ and periodized/{snapshot} protocols}
	\label{fig:phi_convergence}
\end{figure}
As the number of inclusions is fixed to $K^3$ for the periodized ensemble, the volume fractions are exact up to voxelation, i.e., the rastering process which furnishes each voxel either by matrix or filler material. The latter process, as it is based on the voxel's centroid, shows a $L^{-3/2}$-scaling in accordance with the CLT. For the {snapshot} ensemble, however, we observe an $L^{-1}$-scaling. In addition, we also sampled the number of spherical inclusions from a Poisson process, and used the MCM to generate a corresponding periodized Poisson ensemble. Unfortunately, if $K$ was too small, the number of cells produced by the Poisson process often led to a filler content that could not be realized by MCM. For that reason, only results for $K\ge 4$ are shown. Fig.~\ref{fig:phi_convergence_spheres} shows an $L^{-3/2}$-decay of the standard deviation for the periodized Poisson process. However, the absolute values are on a higher level than for the previously discussed protocols. Please note that the mean volume fractions for all three considered scenarios and all considered $K$ were within $0.1\%$ relative error of the target $30\%$ volume fraction.\\
For comparison, the decay rates of the sampled volume fractions of the circular inclusions are shown in Fig.~\ref{fig:phi_convergence_disks}. Both the periodized and the {snapshot-}sampling strategies are characterized by a $1/L$-scaling, in accordance with the CLT.\\
As an intermediate result, we conclude that the non-standard scaling of the random error, which we observed in {Section} \ref{sec:results}, is already present at the level of the volume fraction. Fortunately, understanding the sampling of the volume fraction is simpler than understanding the random error through homogenization. Indeed, there is a (more) direct link between the autocorrelation function and the decay behavior of the volume fraction for the {snapshot} protocol, which arises as the mean of the (random) characteristic-function {snapshot on} cells $Q_L$ of finite size.\\
For estimating the decay rate of the volume fraction for {snapshot} ensembles, the notion of integral range~\cite{EstimatingAndChoosing,GeostatisticalSimulation} is very useful, and we will quickly summarize the important points.\\
Suppose that $Z$ is a stationary random field on $\R^d$ with finite second moment. By stationarity, the mean
\[
	\mu = \mean{Z(x)} \quad \text{{and the variance}} \quad \sigma^2 = \var{Z(x)} \equiv \mean{(Z(x) - \mu)^2}
\]
are independent of $x \in \R^d$. For any compact set $V \subseteq \R^d$, let us denote by $Z_V$ the random variable obtained by averaging $Z$ over $V$
\[
	Z_V = \dashint_V Z(x)\, dx.
\]
$Z_V$ gives rise to an unbiased estimator for the mean $\mu${.}
{For $\sigma>0$,} we are interested in the variance of $Z_V$,
\begin{equation}\label{eq:variance_autocorrelation}
	\var{Z_V} = \mean{ \left( \dashint_V Z(x)\, dx - \mu \right)^2} {\equiv \sigma^2 \, \dashint_V \dashint_V h(x-y)\, dx \,dy},
\end{equation}
{expressed in terms of} the scaled autocorrelation function
\begin{equation}\label{eq:autocorrelation_function}
	h(x) = \sigma^{-2} \, \mean{(Z(0)-\mu)(Z(x)-\mu)}.
\end{equation}
The {identity \eqref{eq:variance_autocorrelation}} uncovers the relation between the variance of the $V$-averaged variable $Z_V$ and the autocorrelation function \eqref{eq:autocorrelation_function}.
If $h \in L^1$, the estimate 
\[
	\var{Z_V} \le \sigma^2 \, \dashint_V \dashint_V |h(x-y)|\, dx \,dy \le \frac{\sigma^2}{|V|} \, \dashint_V \int_{\R^d} |h(x-y)|\, dx \,dy = \frac{\sigma^2 \|h\|_{L^1}}{|V|},
\]
holds, i.e., when evaluated on the cube $V = Q_L$,
\[
	\var{Z_{Q_L}} \le \sigma^2\, \|h\|_{L^1} \, L^{-d}.
\]
In particular, $Z_V$ converges to zero (at least) with the CLT scaling. Thus, if the autocorrelation function $h$ decays to zero sufficiently rapidly, a CLT scaling is ensured. For slower decorrelations, inferior convergence rates may be expected.\\
For the problem at hand, we are interested in $Z = \chi$, the characteristic function of the random inclusions. Then, {we calculate}
\[
	\mu = \phi \quad \text{and} \quad \sigma^2 = \phi(1-\phi)
\]
in terms of the particle volume fraction $\phi$. Also, in terms of the identification $\phi^\text{{sn}}_L = \chi_{Q_L}$, the equation \eqref{eq:variance_autocorrelation} becomes
\begin{equation}\label{eq:normalized_variance_in_the_making}
		\var{\phi^\text{{sn}}_L} = \phi(1-\phi) \, \dashint_{Q_L} \dashint_{Q_L} h(x-y)\, dx \,dy.
\end{equation}
For spherical inclusions, we investigate the autocorrelation function $h$ more closely, cf. Fig.~\ref{fig:autocorrelations_compare}.
\begin{figure}
	\begin{subfigure}{.49\textwidth}
		\includegraphics[width=\textwidth]{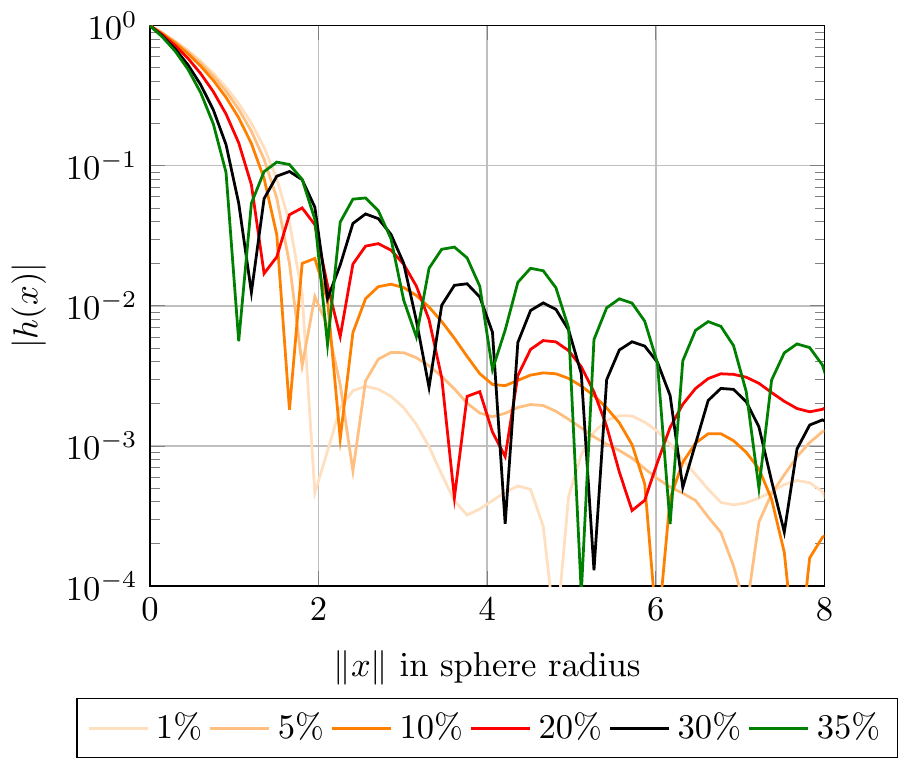}
		\caption{Single realizations at $L/\ell = 32$ and varying fiber volume content $\phi$}
		\label{fig:autocorrelations_compare_variationPhi}
	\end{subfigure}
	\begin{subfigure}{.49\textwidth}
		\includegraphics[width=.945\textwidth]{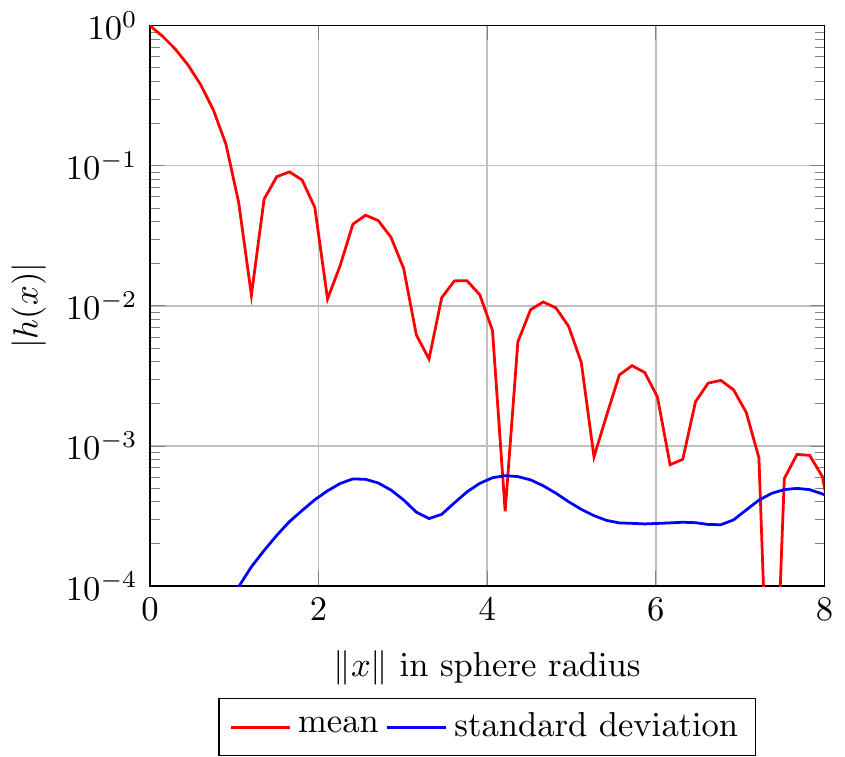}
		\caption{Mean and standard deviation for ten realizations at $L/\ell = 64$ and $\phi = 30\%$}
		\label{fig:autocorrelations_compare_mean_std}
	\end{subfigure}
	\caption{Magnitude of the autocorrelation function $|h|$, depending on $\|x\|$}
	\label{fig:autocorrelations_compare}
\end{figure}
 Indeed, if the autocorrelation function $h$ has a sufficiently rapid decay, the volume fraction $\phi_L$ {of the snapshot} will decay (at least) with the CLT scaling. Conversely, a non-CLT scaling must be rooted in slowly decaying decorrelations.\\
We considered cells with $32^3$ spheres at $\phi=30\%$ as our point of departure, fixed their size and reduced the generated volume fractions. The resulting \emph{empirical} autocorrelation functions, for a single realization and computed by an FFT-based approach~\cite{Fullwood2008}, are shown in Fig.~\ref{fig:autocorrelations_compare_variationPhi}. The autocorrelation function $h$ at $x$ provides the probability of finding an inclusion both at $y$ and $y+x$ (averaged over $y$). As we consider only statistically isotropic ensembles, the autocorrelation function depends only on the magnitude $\|x\|$ of the vector $x$. Thus, the oscillations in Fig.~\ref{fig:autocorrelations_compare_variationPhi} indicate closely packed spheres with a minimum distance between them. This is a result of the enforced minimum distance between the inclusions. With this interpretation in mind, Fig.~\ref{fig:autocorrelations_compare_variationPhi} may be inspected more closely. For increasing volume fraction, the number of "links" in closely packed "chains" of spheres increases. Indeed, even for $\phi=20\%$, only two spheres will be found in the vicinity of a fixed sphere. The third next sphere will be further apart already. For $\phi=30\%$, already seven spheres are found close to a given sphere. For comparison, we also included the $\phi=35\%$ case, which is even more highly packed. Actually, it was not possible to generate $\phi=40\%$ with the MCM method. Indeed, due to the $20\%$ isolation distance, we generate packings of spheres with a larger radius, and down-scale those spheres only in post-processing. For $20\%$ isolation distance and a volume fraction $\phi=40\%$, we would be generating sphere packings without isolation distance and a volume fraction $\phi = 1.2^3 \times 40\% = 69.12\%$. This number exceeds the jamming limit for spheres in three dimensions, and is thus extremely difficult, if not possible, to be realized by a random packing.\\ 
To assess the realiability of the computed autocorrelation functions, we computed ten realizations of the empirical autocorrelation functions for $64^3$ spheres at $30\%$ volume fraction, and considered the mean and standard deviation of these samples, cf. Fig.~\ref{fig:autocorrelations_compare_mean_std}. Notice that the mean autocorrelation function reaches the level of the standard deviation for about ten sphere radii and at a level of about $10^{-3}$. Thus, the actual decay rate of the autocorrelation function {cannot be identified} from \emph{these} numerical investigations.\\
Thus, we turn our attention {to} equation \eqref{eq:normalized_variance_in_the_making}, and assess the normalized quantity
\begin{equation}\label{eq:normalized_standard_deviation}
		\sqrt{\frac{\var{\phi^\text{{sn}}_L}}{\phi(1-\phi)}} =  \sqrt{\dashint_{Q_L} \dashint_{Q_L} h(x-y)\, dx \,dy}.
\end{equation}
directly.
\begin{figure}
	\begin{subfigure}[b]{.49\textwidth}
		\includegraphics[width=\textwidth]{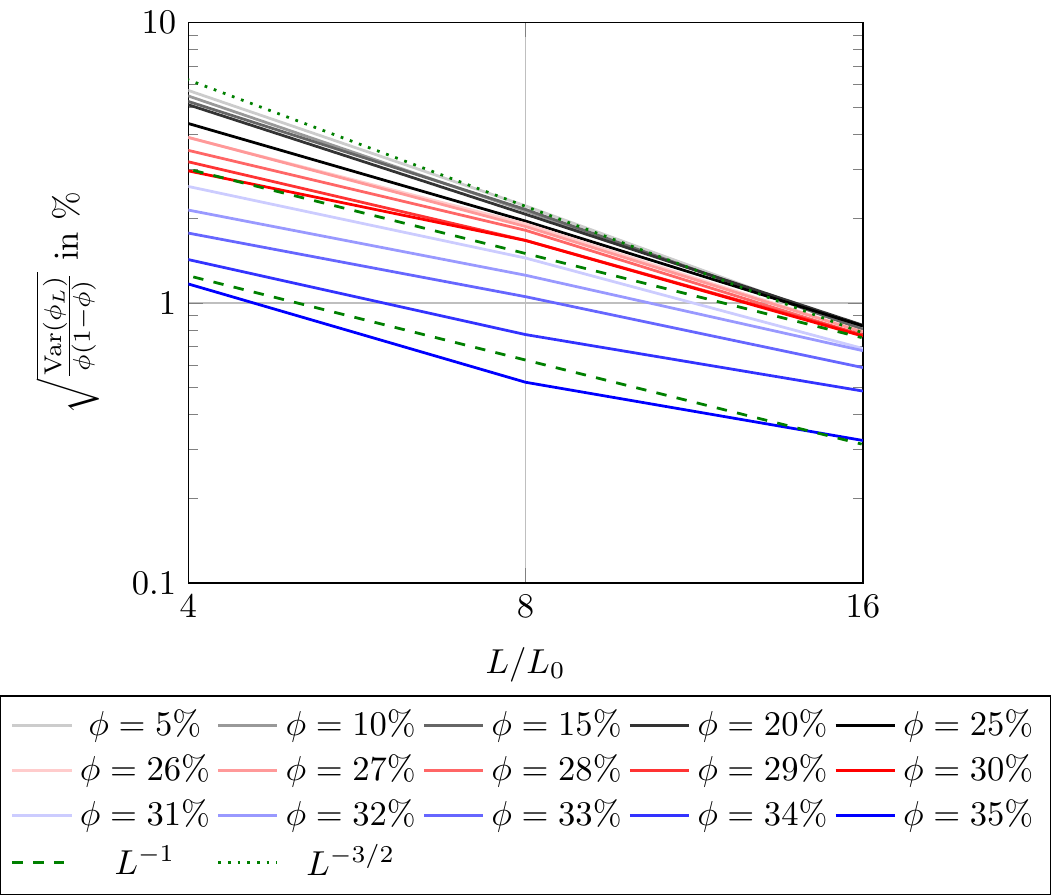}
		\caption{Volume fraction}
		\label{fig:sphere_vf_variation_vf}
	\end{subfigure}
	\begin{subfigure}[b]{.49\textwidth}
		\includegraphics[width=\textwidth]{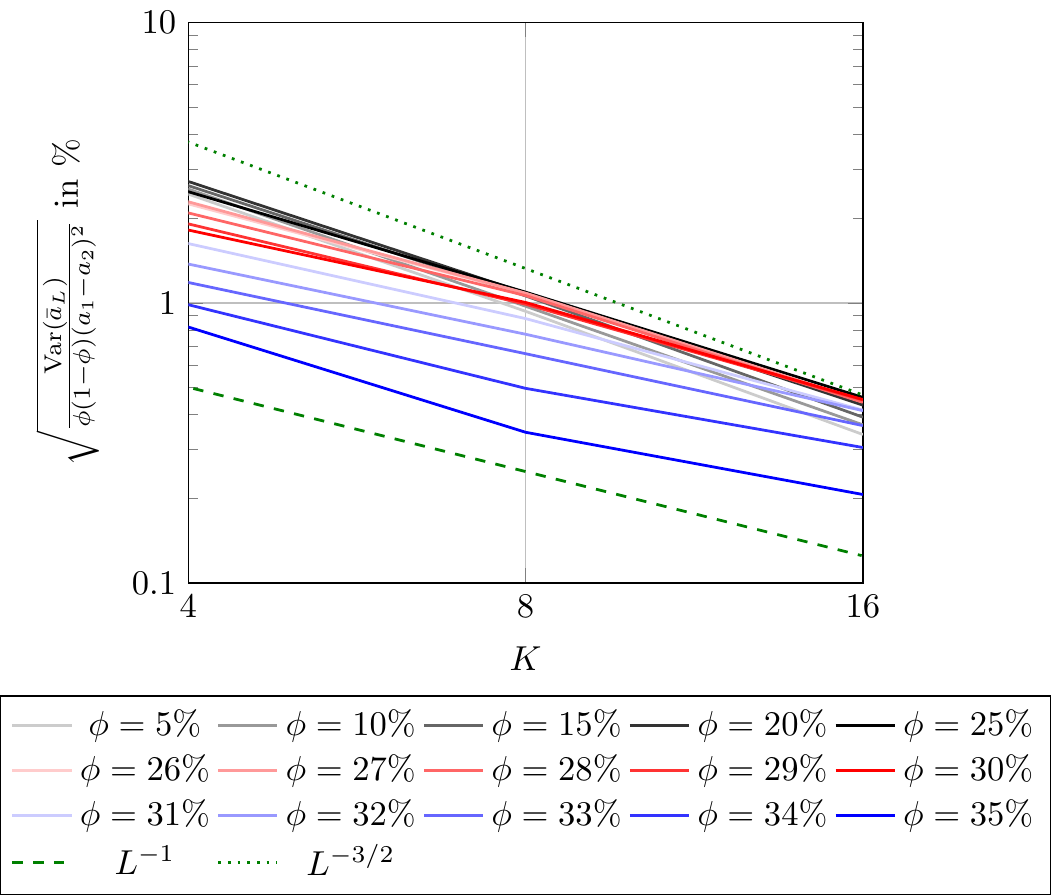}
		\caption{Effective conductivity}
		\label{fig:sphere_vf_variation_ahom}
	\end{subfigure}	
	\caption{Normalized standard deviation of the volume fraction and the effective conductivity for spherical particles}
	\label{fig:sphere_vf_variation}
\end{figure}
For increasing volume fraction, the normalized standard deviation \eqref{eq:normalized_standard_deviation} is shown in Fig.~\ref{fig:sphere_vf_variation_vf}{, where we introduced a reference length scale $L_0$, s.t. at $\phi=30\%$, exactly $(L/L_0)^d$ spheres fit inside a cell with volume $L^d$ to realize this volume fraction. Thus, $L/L_0$ is identical to the parameter $K$ used previously, but also makes sense for other values of the filler fraction $\phi$}. Please note that we only ran \numprint{1000} simulations to cover the great variety of considered volume fractions.\\
Interestingly, up to a volume fraction of $20\%$, the normalized standard deviation is independent of the volume fraction and decays as predicted by the CLT. Thus, we observe a regime of rapid decorrelation. Increasing the volume fraction further, in particular from $25\%$ to $30\%$, leads to an incremental change in the convergence rate, continuously changing from the CLT scaling to a $1/L$-rate. In {Section} \ref{sec:results_spheres}, we considered precisely this regime. Increasing the volume fraction even further retains the $1/L$-rate, but leads to an overall decrease of the standard deviation. Thus, we observe a regime with slow decorrelation but decreased variance due to an increased degree of spatial order.\\
The corresponding empirically computed, normalized standard deviation of the effective conductivity is shown in Fig.~\ref{fig:sphere_vf_variation_ahom}. Similar to the volume fraction, different regimes emerge, corresponding to rapidly decorrelating ensembles which a favorable convergence rate (slightly inferior than CLT) and ensembles with long-range order, and an associated $1/L$-decay.\\
Motivated by these observations, we turn to the two-dimensional case of circular inclusions.
\begin{figure}
	\begin{center}
		\includegraphics[width=.5\textwidth]{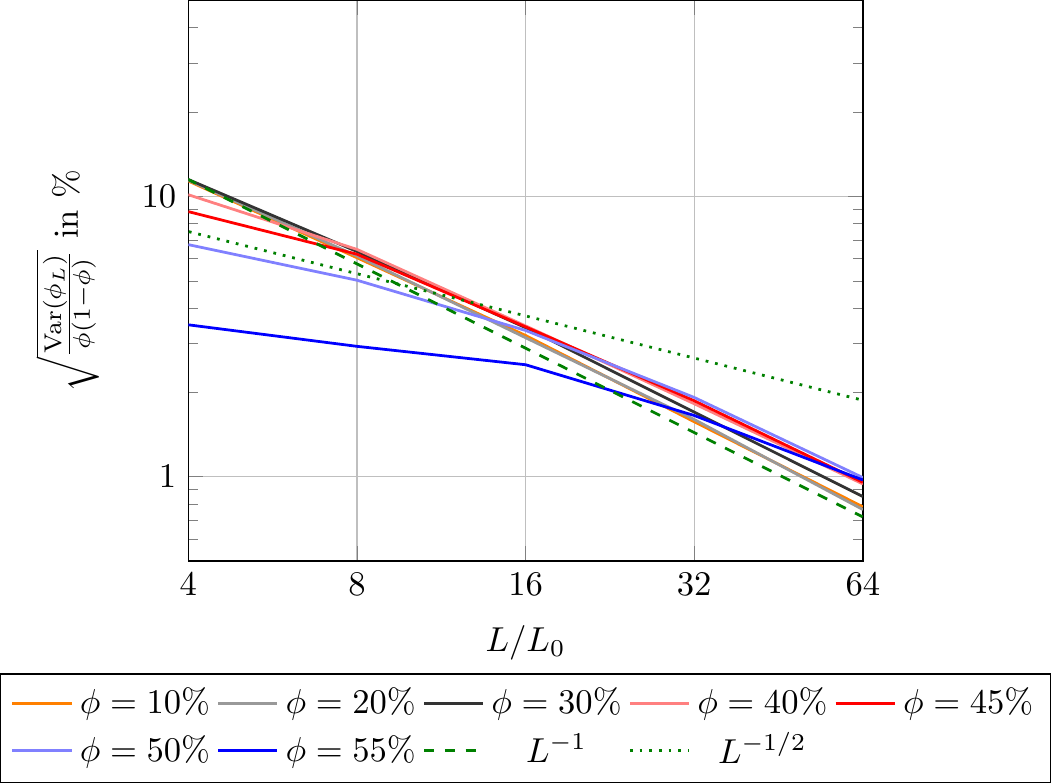} 
	\end{center}
	\caption{Normalized standard deviation of the {area} fraction for circular inclusions}
	\label{fig:disks_vf_variation}
\end{figure}
For spherical inclusions, the observed decay rate deteriorates close to the maximum packing fraction (accounting for the minimum distance). Thus, we similarly increased the packing area fraction for the circular inclusions, cf. Fig.~\ref{fig:disks_vf_variation}. Using {the} MCM algorithm, {area fractions} up to $\phi=55\%$ could be reached. Thus, in contrast to the three-dimensional case, the {area fraction} $\phi=30\%$ considered in {Section} \ref{sec:results_disks} is not even close to the maximum packing fraction. Thus, the corresponding normalized standard deviation of the {snapshot} {area}-fractions decay with the CLT scaling. Only for much higher {area} fractions, at about $\phi=50\%$, an inferior decay rate of $L^{-\frac{1}{2}}$ appears. Interestingly, in such cases, there is a transition from a $L^{-\frac{1}{2}}$-scaling up to $K=16$ to the expected $L^{-1}$-scaling for larger cells.\\
It appears reasonable that such a transition to the standard CLT scaling should also appear for spherical and cylindrical fillers in three spatial dimension. However, the necessary size of the volume elements exceeds current computational capacities. Hence, we cannot observe it in this study.\\
These guesses on a CLT scaling for sufficiently large cells can also be backed up by theory. Indeed, Jeulin ~\cite{Jeulin2016} showed that codimension-$k$ linear varieties randomly dispersed in $\R^{d}$ may lead to {an} $L^{{d}-k}$-scaling of the variance. Willot~\cite{Willot2017} demonstrated that, for cylindrical fibers at low volume fraction, two scaling regimes for the variance of the volume fraction are observed. Indeed, for small cells, the fibers appear infinitely long, and {an} $L^{-1}$-scaling arises. For sufficiently large cells, larger than about ten ~\cite[Sec. 4.3]{Willot2017} times the fiber length, the fibers are "small" compared to the volume, and the expected CLT scaling is recovered.\\
For the spherical and the circular fillers, a similar behavior is observed. Indeed, due to the microstructure generation process, the individual fillers interact in chains, or even networks of {chains}, cf. Fig.~\ref{fig:MCM}. Thus, one-dimensional structures arise. The average length of these structures depends on the target volume fraction, and, in turn, determines the decay behavior, cf. Fig.~\ref{fig:sphere_vf_variation_vf}.
\begin{figure}
	\begin{subfigure}{.49\textwidth}
		\includegraphics[width=\textwidth, trim = 350 130 350 130, clip]{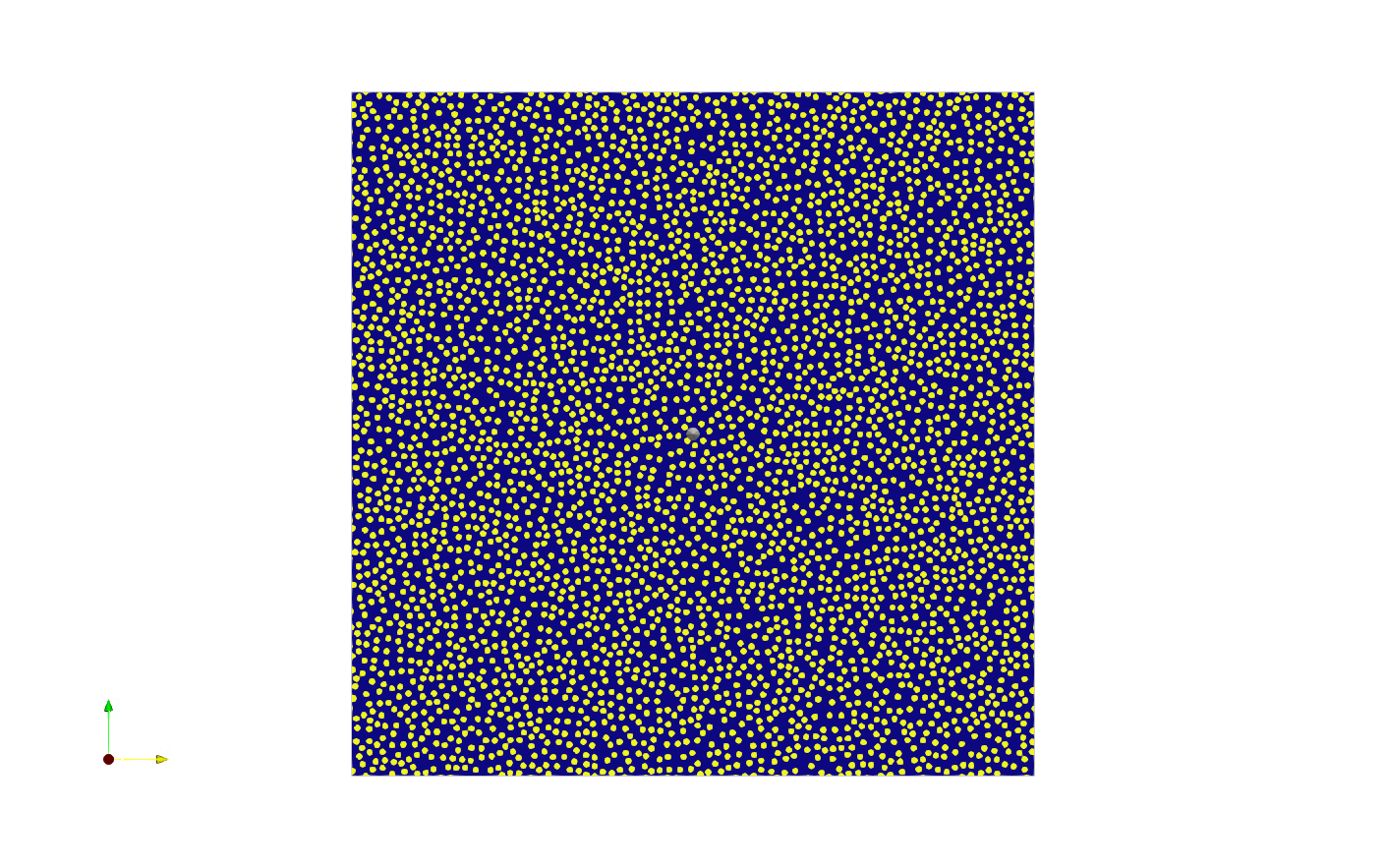}
		\caption{{Random sequential addition (RSA)~\cite{Feder1980}}}
		\label{fig:comparingGenerationMethods_RSA}
	\end{subfigure}
	\begin{subfigure}{.49\textwidth}
		\includegraphics[width=\textwidth, trim = 350 130 350 130, clip]{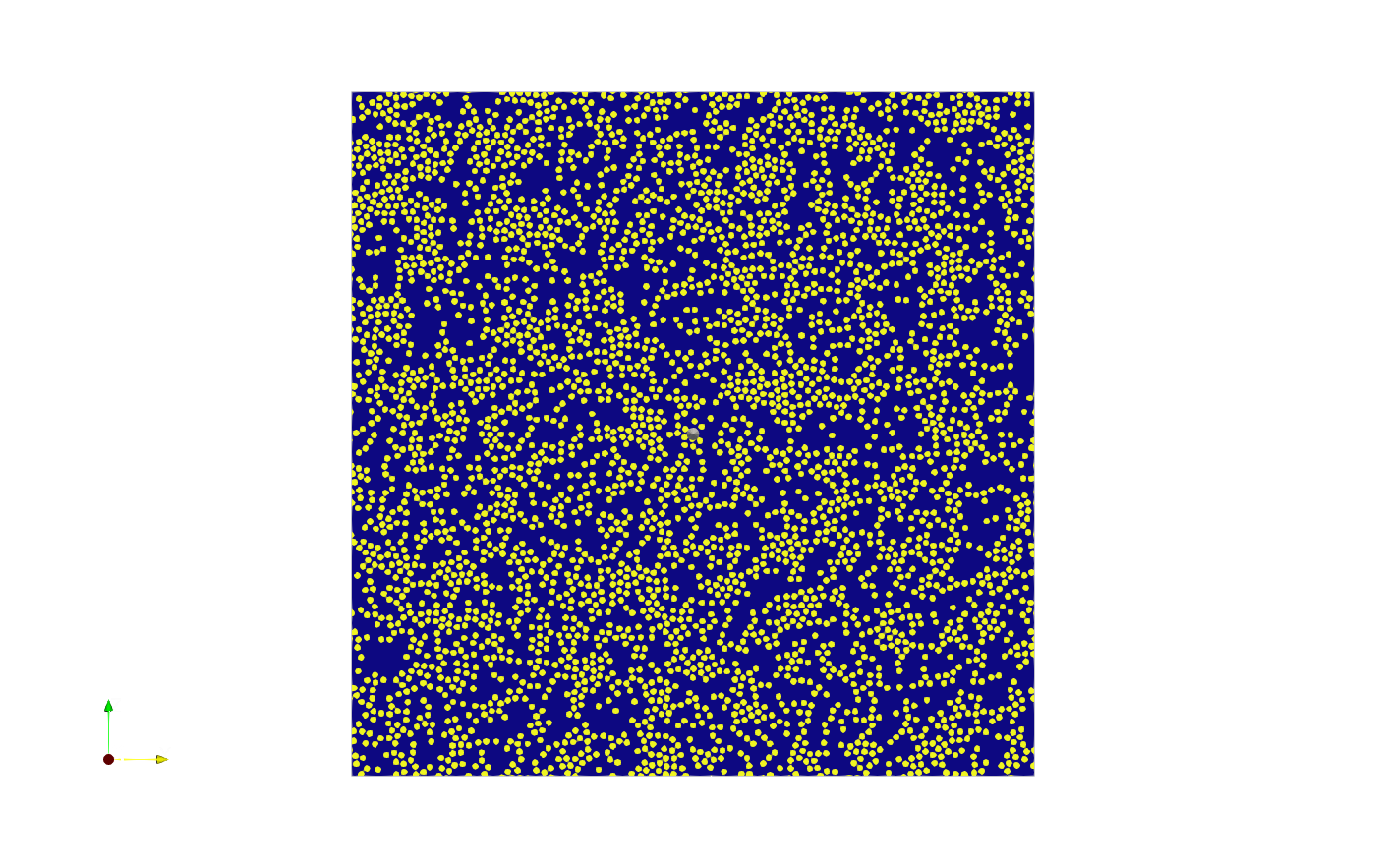}
		\caption{{Mechanical contraction method (MCM)~\cite{WilliamsPhilipse}}}
		\label{fig:comparingGenerationMethods_MCM}
	\end{subfigure}
	\caption{{Visual comparison of two cells with $64^2 = 4096$ circular inclusions at $\phi=30\%$ area fraction, generated by two different algorithms}}
	\label{fig:comparingGenerationMethods}
\end{figure}
{The sitation is visualized in Fig.~\ref{fig:comparingGenerationMethods}. We compare two microstructures with the same microstructure parameters, but generated by different algorithms. The vanilla RSA method, see Fig.~\ref{fig:comparingGenerationMethods_RSA}, leads to a configuration with a much more homogeneous appearance. In contrast, the MCM tends to form clusters, see Fig.~\ref{fig:comparingGenerationMethods_MCM}. This does not come unexpected. Please recall that there is a well-defined jamming limit for RSA, which limits the expected volume fraction to be reached. At this jamming limit, the microstructures attain a rather homogeneous appearance, essentially due to a lack of non-occupied spots.\\
In contrast, the MCM is a \emph{collective re-arrangement} algorithm which permits reaching much larger volume fractions than the RSA. It will also reach some sort of uniform jamming state, but at a much higher filler content.\\
Returning to the \emph{real} material, of course, we cannot ab initio decide which microstructure generation method is suitable for the microstructured material at hand. However, for matrix-inclusion composites with industrial filler content, we expect collective re-arrangement algorithms to be more appropriate. Indeed, sequential algorithms like RSA place the inclusions in a successive fashion. In particular, the correlation length of the emerging ensemble is on the order of the inclusion size, independent of the filler fraction. At high filler content, and also motivated by the real production processes, the individual particles are expected to actually affect their neighbors. Indeed, for instance during injection molding, the filler particles interact with each other, both, in terms of hydrodynamic interactions, and in terms of direct particle-particle collisions. Thus, the correlation length of the produced microstructured materials actually depends on the filler content, and may cover a few particle sizes for high volume fraction.\\
To sum up, the observed inferior decay of the random error for the cut microstructures is a consequence of slowly decaying correlations between the particles in the ensemble. The latter are expected to emerge at high filler content, independently of the microstructure generation algorithm, used. Still, homogenization theory applies as long as the considered volume elements are large compared to the correlation length. Furthermore, in three dimensions and for the considered ensembles, we were unable to work on sufficiently large volume elements to reach the regime of the classical CLT scaling.}

\section{Conclusion}
\label{sec:conclusion}

This work was devoted to sampling strategies for representative volume elements of matrix-inclusion composites.\\
From an engineering point of view, this article may be viewed as a natural continuation of the investigations of Kanit et al.~\cite{KanitRVE}, who introduced statistical terminology and methods for estimating a suitable RVE size. In contrast to Kanit et al.~\cite{KanitRVE}, who studied Voronoi tessellations with seeds distributed according to a Poisson point process, we were concerned with matrix-inclusion composites, i.e., we investigated circular, spherical and cylindrical fillers {embedded} in polymer matrices. In the engineering literature, attention is typically paid to the type of boundary conditions used for the corrector field on the RVE, i.e., whether Dirichlet, Neumann, periodic or mixed boundary conditions are imposed at the cell boundary~\cite{KanitRVE}. For the article at hand, we focused on the treatment of particles at the cell boundary. Indeed, digital volume images of microstructures, which are available at high resolutions, see Bargmann et al.~\cite{BargmannReview} for an overview for different types of microstructures, extract a sample from a component typically by either cutting or piercing {- which we call taking a snapshot in the manuscript at hand}. In particular, the resulting microstructures will not be periodic, and particles at the cell boundary are treated "improperly"{, even \review{when imposing} periodic boundary conditions}. We could quantify the error introduced by {taking snapshots} to be rather high, both in terms of the systematic \emph{and} the random error. The next logical step would be to investigate appropriate measures to counter-balance this effect, for instance by statistical methods~\cite{OhserMuecklich} or modified correctors~\cite{FancyStuff1,FancyStuff2}. Also, the investigations underlined the importance of proper microstructure-generation tools~\cite{BargmannReview}, \emph{even if digital images are available}. Put differently, it is strongly recommended to periodize the {ensemble}, and not only the {snapshot}.\\
In view of the work of Kanit et al.~\cite{KanitRVE}, we could confirm their {observation} that the systematic error for periodized {ensembles} does not exceed the random error, at least for the material contrast considered, {so} that the {random} error is a good indicator for the {total} RVE error.\\
We chose to work with thermal conductivity and a small material contrast to show the high probabilities for failing to produce accurate predictions of effective quantities, emphasizing that already for this {seemingly} innocent scenario, {using the naive snapshot strategy} may be {suboptimal}. Investigating higher material contrast and mechanical problems, in particular inelastic problems, is of immediate engineering interest. Groundwork was already laid by Kanit et al.~\cite{KanitRVE}, who reported an inferior convergence rate for the {random} error associated to mechanical problems. Also, Kanit et al.~\cite{KanitRVE} argued that the {random} error scales as the material contrast, which remains to be confirmed by theoretical means.\\
{Moreover}, it would be interesting to investigate what happens for \emph{long} fibers. In this work, we investigated short and continuous\review{, i.e., infinitely long and aligned,} fibers, but did not account for the intermediate regime {expected in the presence} of long fibers. In that case, RVEs \review{with tractable complexity} need to be significantly smaller than the fiber length, \review{so that} constructing periodized ensembles appears difficult.\\
To conclude the engineering perspective, let us highlight the importance of \review{stochastic notions} for working with RVEs in practice. Typically, emphasis is put on balancing computational effort and accuracy. For instance, if optimal scaling in the degrees of freedom is {desired}, one could either compute on eight cells of size $Q_L$ or a \emph{single} cell of size $Q_{2L}$ in three spatial dimensions. However, there is an often neglected advantage of working with smaller cells - they provide an error estimator free of extra charge. Indeed, based on the empirical variance, in addition to the computed mean value, also a confidence interval may be estimated. Assuming the systematic error to be small compared to the {random} error, the latter confidence interval also provides an estimate for the error relative to the effective property of the infinite medium. In contrast, using only a single computation on the larger cell does not permit such an error analysis.\\
From the viewpoint of mathematical analysis, this article is a natural continuation of previous work by Khoromskaia et al.~\cite{Khoromskiis2019,Khoromskiis2020}, where quantitative homogenization results for random media were confirmed by numerical simulations. Indeed, we go beyond the cited work by investigating microstructure models of higher complexity, as is required for engineering applications. In particular, we break into terrain previously uncharted from a theoretical point of view. {The ensemble which we studied differs from the {simpler} {one} of independently drawing particles conditioned on no overlap. The former permits to reach higher volume fraction than the latter, in general, which is a desirable property for engineering applications. As the studied ensemble {reveals a cross-over in the error scaling at} a volume-fraction dependent intermediate length scale, it appears interesting to investigate this ensemble in more detail{.}}\\
As quantitative stochastic homogenization crucially relies upon quantitative versions of ergodicity, typically in the form of functional inequalities in probability, it may be of interest to study which conditions hold for the ensembles considered in this work, see Duerinckx-Gloria~\cite{WeightedFunctionalInequalities,WeightedFunctionalInequalities2} for investigations covering Poisson-Voronoi tessellations, as treated by Kanit et al.~\cite{KanitRVE}, and random sequential adsorption. In a similar direction, it might be interesting to study the intermediate {length scale below which} the convergence rate of the {random} error {is inferior} for {snapshots of} matrix-inclusion composites.\\
{M}odified corrector problems~\cite{GloriaHabibi2016,FancyStuff1} aim to {upgrade} the decay of the systematic error {to the level of the random error}, but typically assume a CLT scaling of the random error. For the studied matrix-inclusion composites and the {snapshot} protocol, a strictly inferior decay of the random error was observed {in an intermediate regime} for high filler fraction and \review{moderate} cell sizes. It appears sensible to investigate whether {those methods \review{that} modify the corrector equation~\cite{GloriaHabibi2016,FancyStuff2} improve} the decay of the {random} error {for such ensembles}, as well.\\
{As opposed to Khoromskaia et al.~\cite{Khoromskiis2019}}, we could not (yet) confirm the optimal convergence rate of the systematic error and properly periodized ensembles predicted by Clozeau et al.~\cite{CloJoOtXu_2020}. {Computations on even larger cells could reveal the proper decay rate of the systematic error.}
\\
{A cornerstone of the seminal paper of Kanit et al.~\cite{KanitRVE} is the} concept of integral range in spatial statistics~\cite{GeostatisticalSimulation,EstimatingAndChoosing}{, which quantifies the constant in front of the CLT scaling of a second-order stationary random field. In their study of the homogenization commutator, Duerinckx et al.~\cite{QTensor} {related} the} $Q$-tensor{, introduced by Mourrat and coworkers~\cite{GuMourrat2016,MourratOtto2016},} to the stochastic fluctuations {of the apparent properties associated to RVEs. Due to the characterization~\cite[Thm. 2]{QTensor} in terms of a limiting procedure, the $Q$-tensor may be regarded as an extension of the integral range to stochastic homogenization, and a closer investigation appears {desirable}.}\\
Last but not least, let us remark that it would be interesting to understand what happens at interfaces between different heterogeneous materials~\cite{JosienRaithel}, for instance in recent technological lightweight applications where continuously and discontinuously reinforced plastics are combined~\cite{Gorthofer2019}.

\section*{Acknowledgments}
M. Schneider was supported by the German Research Foundation (DFG) within the International Research Training Group "Integrated engineering of continuous-discontinuous long fiber reinforced polymer structures" (GRK 2078). M. Schneider thanks S. Gajek for computational support and H. Andrä for stimulating discussions.\\
This work was initiated while M. Josien held a post-doc position in the Max Planck Institute for Mathematics in the Sciences (MPI MiS), Leipzig.

\appendix
\section{A series expansion for the effective thermal conductivity}
\label{sec:theory_series}

In this appendix, we collect the arguments leading to the {estimate} \eqref{eq:stochastic_error_two_phase_isotropic}
\[
	{\sqrt{\mean{\left\|\bar{A} - A_L\right\|^2}} \le 4 \sqrt{\alpha_1 \alpha_2} \sqrt{\var{\phi_L}} \, \rho + O(\rho^2)}
\]
for the convenience of the reader. Let us fix some positive definite reference thermal conductivity tensor $A^0$. By algebraic manipulations, see  Milton~\cite[Sec. 14.9]{Milton2002}, for fixed $\bar{\xi} \in \R^d$, the random vector field $\xi$ solves the equations \eqref{eq:xi_compatible_stochastic} and \eqref{eq:div_q_stochastic} precisely if $p = (A+A^0)\xi$ solves the Eyre-Milton equation~\cite{EyreMilton1999}
\begin{equation}\label{eq:EyreMiltonEquation}
	(\Id - YZ^0) p = 2 A^0 \bar{\xi}
\end{equation}
involving the Helmholtz reflection $Y = \Id - 2\Gamma$ and the Cayley mapping $Z^0 = (A - A^0)(A + A^0)^{-1}$. As the reflection $Y$ is orthogonal and $Z^0$ is an $L^2$-contraction provided the coefficient field $A$ is essentially bounded and uniformly positive definite~\cite{EyreMilton1999}, the polarization $p$ solving the Eyre-Milton equation \eqref{eq:EyreMiltonEquation} may be represented in terms of a Neumann series
\begin{equation}\label{eq:EyreMiltonPowerSeries}
	p = 2 \sum_{k=0}^\infty \left( YZ^0\right)^k A^0 \bar{\xi}
\end{equation}
with corresponding expectation
\[
	\mean{p} = 2 \sum_{k=0}^\infty \mean{\left( YZ^0\right)^k A^0 \bar{\xi}}.
\]
Taking into account $p = q + A^0\xi$, we obtain
\[
	\bar{A}\bar{\xi} + A^0 \bar{\xi} = 2A^0\bar{\xi} + 2 \mean{Y Z^0 A^0 \bar{\xi}} + 2 \sum_{k=2}^\infty \mean{\left( YZ^0\right)^k A^0 \bar{\xi}},
\]
i.e., using that the expectation vanishes on the image of the Helmholtz projector $\Gamma$,
\begin{equation}\label{eq:EyreMiltonPowerSeries2}
	\bar{A}\bar{\xi} = A^0\bar{\xi} + 2 \mean{ Z^0 A^0 \bar{\xi}} + 2 \sum_{k=2}^\infty \mean{\left( YZ^0\right)^k A^0 \bar{\xi}},
\end{equation}
To proceed, we restrict our attention to two-phase isotropic composites, i.e., we suppose that the thermal conductivity tensor takes the form
\begin{equation}\label{eq:two_phase_isotropic_composite}
	A = \alpha_1 \chi\, \Id + \alpha_2 (1-\chi)\, \Id
\end{equation}
in terms of non-vanishing isotropic heat conductivities $\alpha_1$ as well as $\alpha_2$ and the indicator function $\chi$ of a random set. Define $A^0 = \alpha_0 \, \Id$ with $\alpha_0 = \sqrt{\alpha_1 \alpha_2}$. Then,
\[
	Z^0 = (A - A^0)(A+A^0)^{-1} = \frac{\alpha_1 - \alpha_0}{\alpha_1 + \alpha_0} \chi\, \Id + \frac{\alpha_2 - \alpha_0}{\alpha_2 + \alpha_0} (1-\chi)\, \Id
	= \frac{\sqrt{\alpha_1} - \sqrt{\alpha_2}}{\sqrt{\alpha_1} + \sqrt{\alpha_2}} \left[ \chi\, - (1 - \chi)\right]\Id,
\]
i.e.,
\[
	Z^0 = \rho (2\chi - 1) \Id \quad \text{with} \quad \rho = 
	{ \frac{ \sqrt{\alpha_1} - \sqrt{\alpha_2}  }{ \sqrt{\alpha_1} + \sqrt{\alpha_2} }  }
	\in (-1,1).
\]
Inserting the latter expression into the Neumann series \eqref{eq:EyreMiltonPowerSeries2}, we obtain
\[
	\bar{A}\bar{\xi} = \sqrt{\alpha_1 \alpha_2}\bar{\xi} + 2 \sqrt{\alpha_1 \alpha_2} \rho ( 2\mean{\chi} - 1 ) \bar{\xi} + 2 \sqrt{\alpha_1 \alpha_2} \sum_{k=2}^\infty \mean{\left( YZ^0\right)^k \bar{\xi}},
\]
which we might also write in the form
\begin{equation*}
	\bar{A}\bar{\xi} = \sqrt{\alpha_1 \alpha_2}(1 - 2\rho)\bar{\xi} + 4 \sqrt{\alpha_1 \alpha_2} \rho \mean{\chi} \bar{\xi} + O(\rho^2).
\end{equation*}
We recover the known result, see Milton~\cite[Ch. 14]{Milton2002}, that, to first order in $\rho$, the volume fraction $\phi \equiv \mean{\chi}$ determines the effective conductivity.\\
For the {snapshot} ensemble on the cell $Q_L$, an analogous argument yields the representation
\begin{equation*}
	\bar{A}^{{\textrm{sn}}}_L = \sqrt{\alpha_1 \alpha_2}(1 - 2\rho)\bar{\xi} + 4 \sqrt{\alpha_1 \alpha_2} \rho \phi^\text{{sn}}_L \bar{\xi} + O(\rho^2),
\end{equation*}
where the constant in the Landau-$O$ depends on $L$ and the {snapshot} volume-fraction $\phi^\text{{sn}}_L$ is a random variable determined by
\[
	\phi^\text{{sn}}_L = \dashint_{Q_L} \chi \, dx.
\]
The {snapshot} volume-fraction $\phi^\text{{sn}}_L$ leads to an unbiased estimator for the volume fraction $\phi$, i.e., 
\[
	\mean{\phi^\text{{sn}}_L} = \phi
\]
holds. In particular, the systematic error $\left\|\bar{A} - \bar{A}^\text{{sn}}_L\right\|$ is entirely determined by the $\rho^2$-term.
In contrast, the formula for the difference
\[
	\bar{A} - \bar{A}^\text{{sn}}_L = 4 \sqrt{\alpha_1 \alpha_2} \left(\phi - \phi^\text{{sn}}_L\right) \rho \, \Id{} + O(\rho^2)
\]
permits {bounding} the {random} error in the form
\begin{equation}\label{eq:stochastic_error_two_phase_isotropic_}
	{\sqrt{\mean{\left\|\bar{A} - \bar{A}^\text{{sn}}_L\right\|^2}} \le 4 \sqrt{\alpha_1 \alpha_2} \sqrt{\var{\phi^\text{{sn}}_L}} \, \rho + O(\rho^2).}
\end{equation}
In particular, the {random} error is, up to first order in the material contrast dependent parameter $\rho$, determined by the standard deviation of the {snapshot} volume-fraction $\phi^\text{{sn}}_L$. We close this {Section} with a remark:
	{The expansion \eqref{eq:stochastic_error_two_phase_isotropic} also holds in case of the} periodized ensemble $A^\text{per}_L$. {If the latter} is constructed in such a way that the associated volume fraction for specific $L$ is such that $\phi^\text{per}_L = \phi$ holds, the linear term in $\rho$ {will} vanish for the standard-deviation expansion \eqref{eq:stochastic_error_two_phase_isotropic}, i.e., the random error may be expected to become smaller. The latter idea has been extended to higher-order terms by Le Bris et al.~\cite{SQSpaper} and analyzed by Fischer~\cite{SQFischer}.

\bibliographystyle{ieeetr}
{\footnotesize\bibliography{ms}}

\end{document}